\documentclass[twocolumn,aps,superscriptaddress]{revtex4-1}
\PassOptionsToPackage{usenames,dvipsnames}{color}
\usepackage{amsmath,amsfonts,amssymb,amsthm,graphicx,bbm,enumerate,times,color,mathrsfs}
\usepackage{dsfont}
\usepackage{times}
\usepackage[normalem]{ulem}
\usepackage{mathtools}
\usepackage{lipsum}

 \definecolor{jens}{rgb}{.2,0.7,.9}

\newtheorem{lemma}{Lemma}

\definecolor{martin}{rgb}{0,.4,1}

\definecolor{henrik}{rgb}{1,.4,0}

\newcommand{\mc}[1]{\mathcal{#1}}
\newcommand{\mr}[1]{\mathrm{#1}}

\newcommand{\e}{\mathrm{e}}

\newcommand{\tr}{\mathrm{Tr}} 
\newcommand{\Tr}{\mathrm{Tr}} 

\newcommand{\id}{\mathbb{I}}

\renewcommand{\mod}{\mathrm{\, mod\, }}

\newcommand{\one}{\mathbf{1}}
\newcommand{\1}{\mathrm{id}}
\DeclareMathOperator{\var}{\mathrm{var}}

\newcommand{\norm}[1]{\left\Vert #1 \right\Vert}

\newcommand{\ket}[1]{\left.\left|{#1}\right.\right\rangle}

\newcommand{\bra}[1]{\left.\left\langle{#1}\right.\right|}

\newcommand{\average}[1]{\left\langle #1\right\rangle}

\newcommand{\dd}{\textrm{d}}

\newcommand{\iu}{\textrm{i}}

\newcommand{\fu}{Dahlem Center for Complex Quantum Systems, Freie Universit{\"a}t Berlin, 14195 Berlin, Germany}

\usepackage{hyperref}
\begin{document}
\title{Strong coupling corrections  in quantum thermodynamics}

\author{M.\ Perarnau-Llobet}
\affiliation{Max-Planck-Institut f\"ur Quantenoptik, D-85748 Garching, Germany}
\affiliation{ICFO-Institut de Ciencies Fotoniques, The Barcelona Institute of Science and Technology, 08860 Castelldefels, Barcelona, Spain}
\author{H.\ Wilming}
\affiliation{\fu}
\author{A.\ Riera}
\affiliation{ICFO-Institut de Ciencies Fotoniques, The Barcelona Institute of Science and Technology, 08860 Castelldefels, Barcelona, Spain}
\author{R.\ Gallego}
\affiliation{\fu}
\author{J.\ Eisert}
\affiliation{\fu}

\begin{abstract}
Quantum systems strongly coupled to many-body systems equilibrate to the reduced state of a global thermal state, deviating from the local thermal state of the system as it occurs in the weak-coupling limit.  Taking this insight as a starting point, we study the thermodynamics of systems strongly coupled to thermal baths. First, we provide 
strong-coupling corrections to the second law applicable to general systems in three of its different readings: As a statement of maximal extractable work, on heat dissipation, and bound to the Carnot efficiency. These corrections  become relevant for small quantum systems and always vanish in first order in the interaction strength. We then move to the question of power of heat engines, obtaining a bound on the power enhancement due to strong coupling. Our results are exemplified on the paradigmatic situation of non-Markovian quantum Brownian motion.
\end{abstract}
\maketitle

\section{Introduction}

Thermodynamics is the fundamental theory
 concerned with heat and temperature and their relation to energy and work. 
 In phenomenological  thermodynamics, an implicit assumption is
 that couplings between
the working systems and their heat baths are so weak so that effects of the interaction can be
neglected. As a consequence, the equilibrium states of the working systems are thermal states, in
fact thermal states of local Hamiltonians.
For small-scale systems governed by quantum mechanical laws, however, such a
weak-coupling limit can be far from being reasonable,
as the surface area of such systems is often not much smaller than
their volume. An impressive body of literature in a related field, namely
equilibration and thermalisation of closed quantum many-body systems
\cite{Gogolin2016},
strongly suggests that a system coupled strongly to a thermal bath
should be described by
the \emph{local reduced state of the global Gibbs state}
$\rho_S = \Tr_B (e^{-\beta H}/{\rm tr}(e^{-\beta H}))$
-- and not by a Gibbs
state of the local Hamiltonian itself \cite{Subasi2012,Mueller,BrandaoCramer,Farrelly16}.

In this work we take this basic but profound insight seriously when studying
in detail quantum thermal machines strongly coupled to heat baths.
First, we prove exact and general bounds
on work extraction from a non-equilibrium system that can be brought in contact with a single heat bath.
These results can be captured
as universal corrections to the weak coupling limit -- showing that strong coupling
unavoidably leads to irreversibility and is hence detrimental for work
extraction. Similar corrections are obtained for heat dissipation
and the Carnot efficiency, hence providing strong-coupling corrections to
the different formulations of the second law of thermodynamics. For thermal machines, we
also show that strong interactions lead to power enhancements.
Finally, we illustrate these considerations by means of the paradigmatic quantum Brownian motion  \cite{Caldeira1983}.

These results are put into context of the
field of quantum thermodynamics \cite{Vinjanampathy2016,Goold2016,KosloffEnt}, in particular,
of recent efforts to describe the thermodynamics of quantum systems with strong interactions between system and bath  \cite{Gelbwaser2015,Uzdin2016,Kkatz2016,Strasberg2016,Newman2016,Xu2016,Lee2012,Hanggi2008,Gelin2009,Campisi2009,Mahler2010,Campisi2010,Lutz2011,
Lee2012,Gallego14,Wang2015,Esposito2015prl,Esposito2015,Bruch2016,Perarnau-Llobet2015a,Kato2016,Philbin2016,Carrega2016b} (see
Refs.\ \cite{Seifert2016,Jarzynski2017} for classical systems). These include considerations on heat engines \cite{Gelbwaser2015,Uzdin2016,Kkatz2016,Strasberg2016,Newman2016,Xu2016}, equilibrium and non-equilibrium thermodynamics \cite{Hanggi2008,Gelin2009,Campisi2009,Mahler2010,Campisi2010,Lutz2011,
Lee2012,
Gallego14,Wang2015,Esposito2015prl,Esposito2015,Bruch2016,Perarnau-Llobet2015a,Kato2016,Philbin2016,Carrega2016b} and, in a more abstract level, generic limitations on transformations between states using thermal resources \cite{ResourceTheory,SecondLaw,Piotr14,Lostaglio14,Wilming15}. The key contribution of the present work, compared with earlier strong-coupling analyses
of heat engines 
\cite{Gelbwaser2015,Uzdin2016,Kkatz2016,Strasberg2016,Newman2016,Xu2016,Lee2012}, is to 
provide bounds on work and efficiency, without having to restrict to any particular model for the systems involved. Our bounds apply to thermodynamic scenarios in which the system equilibrates to the reduced of a global Gibbs state,
and for which the coupling can be switched on and off. More precisely, our results are derived within a framework applicable to general situations; after all, also phenomenological thermodynamics is widely applicable by largely abstracting from the specifics of a given setting.



\section{Framework}

We consider a system $S$,  a heat
bath $B$, with internal Hamiltonians  $H_S$ and $H_B$, respectively. They can interact via a a possibly strong interaction $V$.
Thermodynamic protocols then consist on transformations over $H_S$, and equilibration processes induced by $V$. Specifically,
we consider protocols of $N$ steps, and denote by $\rho^{(i)}$ and $H^{(i)}$ the state and Hamiltonian of $SB$ in the $i$th step,
consisting of three elementary operations
\begin{itemize}
	\item[(A)]\emph{Turning on/off interaction:} With this, we model the process of bringing $S$ and $B$ into contact, so that the Hamiltonian takes the form $H^{(i)}= H_S^{(i)} +H_B+V$.
Similarly, the interaction can be turned off at any step of the process.
Treating such processes as quenches, the average work gain when placing/removing $V$ is 
\begin{equation}\label{eq:workonoff}
W^{(i)}_{\text{on}}= \tr (\rho^{(i)} V)=-W^{(i)}_{\text{off}}.
\end{equation}

\item[(B)]
\emph{A quench on $S$:} A fast transformation of $H_S$ is implemented, so
that  $H^{(i)}= H_S^{(i)} +H_B+V$ is changed to $H^{(i+1)} =H_S^{(i+1)} +H_B+V,$
whereas the state $\rho^{(i)}$ remains unchanged. The corresponding  work gain reads
\begin{equation}\label{eq:workquench}
W^{(i)}=\Tr (\rho_{S}^{(i)}(H_{S}^{(i)}-H_{S}^{(i+1)} ))
\end{equation}
which depends only on the state of $S$, since the interaction energy and the bath energy remain constant
\footnote{Quenches on $S$ can be similarly applied when the interaction $V$ is turned off. Note that in this case Eq.\ (\ref {eq:workquench}) is also valid.}.

\item[(C)]
\emph{A thermalisation  process.} This operation models the closed free evolution of $SB$  when $V$ is present, i.e.,  under $H^{(i+1)}$. In this case, $S$ and $B$
exchange energy, while the total energy is preserved. Hence, this operation has no work cost.
When they reach
equilibrium, we assume that the state of $S$ is well described
by
\begin{equation}
\rho_S^{(i+1)} = \Tr_B (\omega_{\beta}(H^{(i+1)})),
\label{eqstrong}
\end{equation}
where  $\omega_{\beta}(H)=e^{-\beta H}/{\rm tr}(e^{-\beta H})$. Similarly, we assume that the boundary between $S$ and $B$,
i.e., the support of $V$, can also be described
by the reduced of a global  thermal state.
\end{itemize}
Both assumptions are reasonable for locally interacting systems
and are
backed by a body of rigorous arguments \cite{Gogolin2016,Farrelly16}
\footnote{Indeed, in the specific situation of the system $S$ and its bath $B$ being translationally invariant, Ref.\ \cite {Farrelly16} shows that the reduced state of the infinite time average is close in $1$-norm to $\rho _S = {\protect \rm  tr}_B(e^{-\beta H}/{\protect \rm  tr}(e^{-\beta H}) )$, once one perturbs a Gibbs state by acting with a completely positive map on it supported on a constant number of sites only. Here, in contrast, we consider Hamiltonian quenches in $S$. However, in the high temperature limit \cite {Kliesch2013}, the effect of such quenches can be approximated by completely positive maps acting on $S$ and a constant number of degrees of freedom of $B$. The closeness to the infinite time average is guaranteed for the overwhelming majority of times by invoking the equilibration results of Ref.\ \cite {Linden2012} in conjunction to the bounds on the effective dimension from Ref.\ \cite {Farrelly16}.}.
When it is clear from the context we will use the notation
$\omega^{(i)}:= \omega_{\beta}(H^{(i)})$ and $\omega_{S}^{(i)} :=  \omega_{\beta}(H_{S}^{(i)})$. We also use
the convention $\hbar=1$, $k_B=1$, and that when $SB$ decrease their global energy, then
work is extracted and $W>0$. 

A thermodynamic protocol then consists of an arbitrary sequence of operations of the type (A)-(C). The total expected work $W$ gained in the process is the sum of all the contributions of the form \eqref{eq:workonoff} and \eqref{eq:workquench}.
Note that in this framework, the Hamiltonian terms $V$ and $H_B$ remain fixed
throughout the protocol, reflecting the fact that an experimenter will in many realistic situations not have
precise control over $B$ and the coupling between $S$ and $B$, at least not
beyond the capability of turning it on and off.
After every transformation of the form (C), $S$ is assumed to be brought to equilibrium after sufficiently long time. That is, possible finite-time effects are not included in this framework.

\section{Maximal work extraction for arbitrary coupling strengths.} 

We now study work extraction from an out-of equilibrium state of $S$. 
In order to avoid the possibility of extracting work from the energy stored in $V$, we consider that $S$ is initially isolated from $B$. The initial Hamiltonian is hence non-interacting,
$ H^{(0)} = H_S +H_B$, and the initial state is uncorrelated,
$\rho^{(0)} = \rho_S \otimes \omega_\beta(H_B).$ Given these initial conditions, 
the task is  to optimize the extracted work over all cyclic Hamiltonian processes
under the operations (A)-(C). Cyclicity here means that in a  protocol of $N+1$ steps,   we have $H^{(N +1)} =
H^{(0)}$, where $N$ can be arbitrarily large. 

It is instructive to first recall the optimal protocol in the weak-coupling regime \cite{Alicki2004,Esposito2011,Anders2013,Gallego14}. It consists of four steps: (i) a quench from $H_S^{(0)}$ to $\tilde{H}_{S}$, where $\omega_{\beta}(\tilde{H}_{S})=\rho_S$, (ii) turning on  $V$, (iii)   an isothermal process from $\tilde{H}_{S}$ back to $H_S$, and (iv) turning off $V$.  In our framework, isothermal processes correspond to a concatenation of infinitesimally small quenches followed by equilibration steps -- we refer the reader to 
Refs.\ \cite{Anders2013,Gallego14} for more details. The protocol (i)-(iv) has no dissipation, and is hence reversible, in the limit of an arbitrarily weak $V$.
In the strong-coupling regime, where the energy of $V$ can no longer be neglected, we show in Appendix \ref{Sec:WorkExtraction} that the optimal protocol also has the form (i)-(iv),  but the initial and
final Hamiltonians of the isothermal process need to be modified. Let
$H_S^{(1)}$ and $H_S^{(N)}$ be the Hamiltonians of $S$ when $V$ is turned on and off, respectively. Then, the total work $W$ of the  protocol can  be expressed as 
\begin{equation}
W = W^{\rm (weak)}-\Delta F^{\rm (res)} -\Delta F^{\rm (irr)},
\label{WcomparisonBound}
\end{equation}
where $W^{\rm (weak)}=F(\rho_S,H_S)-F(\omega_\beta(H_S),H_S)$ is the maximal extractable work in the weak coupling regime, $F(\rho,H):=\tr(\rho H)+T \tr(\rho \ln \rho)$ is the (non-equilibrium) free energy,  and 
\begin{align}
\label{DefIrrevmain}
\Delta F^{\rm (irr)} &\coloneqq  F(\rho^{(0)},H^{(1)})-F(\omega^{(1)},H^{(1)}),
\\
\label{DefResidualmain}
\Delta F^{\rm (res)} &\coloneqq  F(\omega^{(N)},H^{(0)})-F(\omega^{(0)},H^{(0)}) ,
\end{align}
with $H^{(1)/(N)} = H_S^{(1)/(N)} + H_B + V$.
It is important to note that $F(\rho,H)-F(\omega_\beta(H),H)=T {\rm S}(\rho\|\omega_\beta(H))\geq 0$
 with ${\rm S}(\rho \| \sigma)\coloneqq \tr \left(\rho (\log \rho-\log \sigma)\right)$ the quantum relative entropy.
It follows that always 
$\Delta F^{\mathrm{(irr)/(res)}}\geq 0$, and  we can already conclude that strong coupling cannot be beneficial for work extraction as $W \leq W^{\mathrm{(weak)}}$. The correcting term $\Delta F^{\rm (irr)}$ can be interpreted as the energy dissipated when $S$ is put in contact with $B$, whereas  $\Delta F^{\rm (res)}$ is  the extractable work left on the final state.

The extracted work $W$ in \eqref{WcomparisonBound} is maximised when  $H_S^{(1)}$ and $H_S^{(N)}$
minimise the correcting terms  $\Delta F^{\rm (irr)}$ and $\Delta F^{\rm (res)}$, respectively. Assuming that $\rho_S$ is a full rank state, we show in Appendix \ref{Sec:WorkExtraction} that this happens for    
\begin{align}
 &\hspace{-1.5mm}\tr_B \left(\omega_{\beta}(H^{(1)})\right)=\rho_S,
 \label{X^1}
\\
&\hspace{-1.5mm}\tr_B\hspace{-0.5mm}\left(\hspace{-0.5mm}\omega_\beta(H^{(N)})\hspace{-0.5mm}\right)\hspace{-0.5mm}= \hspace{-0.5mm}\frac{\hspace{-0.5mm}\tr_B\hspace{-0.5mm}\left(\hspace{-0.5mm}\omega_\beta(H^{(N)}) (H^{(0)}_{H^{(N)},\beta}-H^{(N)})\hspace{-0.5mm}\right)}{\tr\left((H^{(0)}_{H^{(N)},\beta} - H^{(N)}) \omega_\beta(H^{(N)})\right)}\hspace{-0.5mm},
\label{X^N}
\end{align}
where we have defined  
$Y_{H,\beta} := \int_0^{1} \hspace{-1mm}d \tau \hspace{1mm} e^{ \beta \tau H} Y e^{- \tau \beta H}$
for Hermitian operators $Y$ -- an integral that can be solved analytically. Furthermore, $\Delta F^{\rm (irr)/(res)}$  have at least one minimum, so that 
Eqs.\ \eqref{X^1}, \eqref{X^N} always provide the desired solution  (see Appendix \ref{Sec:WorkExtraction}).
 If more than one
minimum exists, the solution  corresponds to the absolute one.

It is important to stress that although a priori \eqref{X^1}, \eqref{X^N}, and in general  the extracted work $W$, depend on the entire bath $B$, commonly its Hamiltonian is local and the correlations between its degrees of freedom
decay rapidly with the distance. Therefore, only the degrees of freedom that are geometrically close to
the boundary between $S$ and $B$ will contribute. This has the important consequence that we can
solve ~\eqref{X^1}, \eqref{X^N} by considering a small buffer region in $B$ involving a few degrees of freedom
only, while maintaining tight bounds for the error made in such a prescription
(see Refs.~\cite{Kliesch2013,Ferraro2012,Hernandez-Santana2015}). 
This renders the solution practically and efficiently computable.

Altogether, our techniques provide a procedure to determine for any model the optimal protocol for  work extraction in the strong coupling regime. Essentially, it consists of an isothermal process, where $S$ is put in contact with $B$ according to Eqs.~\eqref{X^1} and \eqref{X^N}. In what follows, we solve explicitly these equations at lowest  order in the interaction strength. 
 
\section{Corrections at lowest order of work}
Interestingly, in a perturbative treatment,
the problem at hand can be essentially solved by computing covariances.
We start by replacing $V$ by $gV$, where the dimensionless $g>0$ 
quantifies the interaction strength. 
Expanding  Eqs.~\eqref{X^1} and \eqref{X^N}  in $g$, we get
\begin{align}
&H_S^{(1)}=\tilde{H}_S-g\tr_B (\omega(H_B)V)+O(g^2),
\label{X^1g}
\\
&H_S^{(N)}=H_S- g \tr_B (\omega(H_B)V)+O(g^2),
\label{X^Ng}
\end{align}
where we recall that $\tilde{H}_S$ is defined via $\rho_S=\omega_{\beta}(\tilde{H}_S)$. 
Inserting \eqref{X^1g} and \eqref{X^Ng} into \eqref{DefIrrevmain} and \eqref{DefResidualmain} respectively, in Appendix \ref{Sec:WorkExtraction} we obtain  $\Delta F^{\rm (irr)/(res)}_{\rm min}:=\min_{H_S^{(1)/(N)}} \Delta F^{\rm (irr)/(res)}$ at lowest  non-vanishing order in $g$ 
\begin{align}
&\Delta F^{\rm (irr)/(res)}_{\rm min} \hspace{-0.5mm}= \frac{\beta g^2}{2} {\rm cov}_{\omega_{\beta}(\tilde{H}^{(0)}/H^{(0)})}\hspace{-0.5mm}(\tilde{V},\tilde{V})\hspace{-0.3mm}+\hspace{-0.3mm}O(g^3).
\label{DeltaFirrg}
\end{align}
Here, we have defined $\tilde{V}:=V-\Tr_B(V \omega_{\beta}(H_B))$,  $\tilde{H}^{(0)}:= \tilde{H}_S+H_B$, and  
${\rm cov}_{\omega_{\beta}(H)}(A,B)=\tr(A_{H,\beta} B \omega_{\beta}(H))-\tr(A \omega_{\beta}(H))\tr( B \omega_{\beta}(H))$
is the generalized covariance \cite{Kliesch2013}, also known as Kubo-Mori inner product in linear response theory \cite{Mori1956,Kubo1957,Mori1965}. 
Some important remarks are now in order,
\begin{itemize} 
\item The first order correction to $W$  vanishes
for any $H_S^{(1)}=\tilde{H}_S+O(g)$, $H_S^{(N)}=H_S+O(g)$.
This follows from the penalty terms $\Delta F^{\rm (irr)/(res)}$ being
differentiable functions of $g$ and having a minimum at $g=0$. The choice
\eqref{X^1g}, \eqref{X^Ng}  provides the minimum coefficient of
$O(g^2)$.   
\item The first order correction in
\eqref{X^1g}, \eqref{X^Ng} exactly compensates for the term $g\tr_B
(\omega(H_B)V)$  which often appears in open quantum systems  as an effective
action of $B$ on $S$ \cite{Weiss93,Breuer2002}.
\item    The generalised covariance ${\rm
cov}_{\omega_{\beta}(H)}(A,B)$ captures the linear response of the thermal
state under perturbations  \cite{Mori1956,Kubo1957,Mori1965}. 
\end{itemize}

\section{Heat and dissipation}
 Let us now turn to heat dissipation in an isothermal process in the strong coupling regime. For that,  we do not consider a cyclic process, and  we instead fix the intial and final Hamiltonian to be $H_S$ and $H_S^{(N)}$ respectively -- specifically $ H^{(N)} = H_S^{(N)} +V+H_B$ and $ H^{(N+1)} = H_S^{(N)} +H_B$. We consider the same initial state as in the work-extracting protocol,  i.e., $\rho^{(0)} = \rho_S \otimes \omega_\beta(H_B)$.

From the first law of thermodynamics, the total heat reads $Q=\Delta E_S+W$,
with $\Delta E_S=\Tr(H_S^{(N)} \omega^{(N)})-\Tr(H_S\rho^{(0)})$.
Since $\Delta E_S$ is fixed by $\rho_S$ and $H_S^{(N)}$, it
becomes clear that the optimal protocol for maximising $W$ also minimises
dissipation.
Then from \eqref{WcomparisonBound},  we  obtain (see Appendix \ref{Sec:Heat} for details)
\begin{equation}\label{eq:Qin}
Q= T\Delta {\rm S} - (\Delta F^{\mathrm{(res)}}_B+T I(\omega^{(N)};S:B)+\Delta F^{\mathrm{(irr)}})
\end{equation}
where $\Delta {\rm S}={\rm S}(\Tr_B(\omega^{(N)}))-{\rm S}(\rho_S)$ is the gain of entropy of $S$,
$\Delta F^{\mathrm{(res)}}_B=T{\rm S}(\Tr_S(\omega^{(N)})\|\omega_\beta(H_B))$ is the increase of the free energy of $B$, and $I(\omega^{(N)};S:B)>0$ is the mutual information between $S$ and $B$.
Note that in the strong coupling case,
$Q < T\Delta {\rm S}$, even when the isothermal process is accomplished reversibly. Again, this is due to the penalising terms
$\Delta F^{\mathrm{(res)}}_B$
and $\Delta F^{\mathrm{(irr)}}$, in addition to the correlations captured by the mutual information.
Minimising dissipation corresponds to  minimising the negative terms in \eqref{eq:Qin}.
However, in this case $\Delta F^{\mathrm{(res)}}_B$ and $I(\omega^{(N)};S:B)$ are fixed through $H^{(N)}$.  Hence, we only have freedom to minimise $\Delta F^{\mathrm{(irr)}}$, a problem that  has been  solved in \eqref{X^1}.

Similar to the case of work, we can now expand the correcting terms over the interaction strength $g$. As before, the first order correction vanishes, so that the series expansion  reads
\begin{align}
Q = T \, \Delta {\rm S} - K_q g^2+  O(g^3),
\label{correctionsQ}
\end{align}
with $K_q>0$, and where we note that $\Delta {\rm S}$ depends on
$\omega_{\beta}(H^{(N)})$ and hence indirectly also on $g$ \footnote{If one wishes to expand $Q(g)$ with respect to $Q^{\protect \rm  (weak)}$, then the first order corrections do not vanish. This is discussed in more detail in Appendix \ref{Sec:Heat}.}. From \eqref{eq:Qin}, we note that a
simple and useful lower bound for $K_g$ is given by $K_q\geq \Delta F^{\rm (irr)}_{\rm min}$ as given by \eqref{DeltaFirrg}. In
other words, Eq.\ \eqref{DeltaFirrg} also provides a
strong coupling correction to dissipation, and to the Clausius formulation of
the second law.

\section{Heat engines}
  Given \eqref{WcomparisonBound} and \eqref{eq:Qin}, it is straightforward to study the efficiency of a heat engine in the strong coupling regime. We consider engines made up of two baths at different temperatures which can
sequentially
interact strongly with $S$. More precisely, we extend our formalism to account for equilibrations,  always in the form \eqref{eqstrong}, with two baths $B_c$ or $B_h$, at two different (inverse) temperatures, $\beta_c$ and $\beta_h$. The task is then to maximise the efficiency of a cycle of the engine.

Not surprisingly, the optimal cycle turns out to have the same form as a Carnot
engine, as we show in Section  \ref{Sec:CarnotEngines} of the Appendix. The Carnot cycle can be described with four steps
\begin{itemize}
\item  an isothermal transformation in
contact with $B_h$ from $H^{(A)}_{S}$ to  $H^{(B)}_{S}$,
\item  a quench from $H^{(B)}_{S}$  to $H^{(C)}_{S}$,
\item an isothermal transformation with $B_h$ from $H^{(C)}_{S}$ to  $H^{(D)}_{S}$,
\item a quench from $H^{(D)}_{S}$back to $H^{(C)}_{S}$.
\end{itemize}
 In the  weak coupling regime, the efficiency is maximised
through the choice of Hamiltonians
\begin{align}
\omega_{\beta_h}(H_S^{(B)/(A)})=\omega_{\beta_c}(H_S^{(C)/(D)}),
\end{align} 
which guarantees no dissipation. Given our previous results, these conditions
are naturally extended in the strong coupling regime to
\begin{align}
\tr_{B_h}\omega_{\beta_h}(H^{(B)/(A)})=\tr_{B_c}\omega_{\beta_c}(H^{(C)/(D)}),
\end{align}
where $H^{(X)}:=H_S^{(X)}+V+H_B$. This provides a simple recipe for
constructing minimal dissipation engines in the strong coupling regime. The
corresponding (maximal) efficiency  $\eta$,  using $\eta=1-|Q_c|/|Q_h|$,
\eqref{correctionsQ} and expanding in $g$, reads 
\begin{align}
\eta = \eta^{\rm C} - g^2   \frac{T_c}{T_h} \left(\frac{K_q^{(h)}}{Q_h^{\rm ( weak)}}+\frac{K_q^{(c)}}{Q_c^{\rm ( weak)}}\right) + O(g^3),
\label{eq:efficiencywithg}
\end{align}
where $\eta^C$ is the Carnot efficiency, $Q_{h/c}^{\rm ( weak)}=T_{h/c}\Delta S^{\rm ( weak)}$ and the entropy change in the weak-coupling regime is defined as $\Delta S^{\rm ( weak)}=S(\omega_{\beta}(H_S^{(B)})- S(\omega_{\beta}(H_S^{(D)})$ and $K_q^{(h/c)}$ are coefficients obtained from \eqref{correctionsQ}  for $B_{h/c}$ (see details in Appendix \ref{Sec:CarnotEngines}).  By recalling the bound $K_q\geq \Delta F^{\rm (irr)}_{\rm min}$ (at order $O(g^2)$), through \eqref{DeltaFirrg} we obtain strong coupling corrections to the Carnot efficiency.

\section{Macroscopic Limit}
  Let us briefly discuss the macroscopic limit, in
which $S$ becomes large.  The correcting terms to work and heat in
\eqref{WcomparisonBound} and \eqref{eq:Qin} can be bounded by the interaction
strength as 
\begin{align}
&\Delta F^{\rm (res)/(irr)}_{\rm min} \leq 2\norm{V},
\nonumber\\ 
&TI(\omega^{(N)};S:B) \leq 2\norm{V},
\end{align}
where $\norm{V}$ is the operator norm of $V$.
The first bound is derived in Appendix \ref{Sec:WorkExtraction}, whereas the second one follows
from Ref.\ \cite{Wolf2008}.  
Now we use these bounds to provide a simple argument to show that the limiting terms disappear when dealing with large systems. Let $S$ be a locally interacting system made up of $n^3$ particles. Let it be coupled also locally with $B$, so that the number of particles interacting with $B$ is $\alpha n^2$, $\alpha$ being a parameter that depends on the specific geometry of the boundary between $S$ and $B$. Let us write the interaction as
\begin{align}
V= \sum_{j=1}^{\alpha n^2}h_j,
\end{align}   
where $h_j$ contains all interactions with the $j$-th particle of $S$ in the boundary. Now we have that,
\begin{align}
\norm{V} \leq  \sum_{j=1}^{\alpha n^2} \norm{h_j} \leq \alpha n^2 \max_{j} \norm{h_j}. 
\end{align}
On the other hand, the extractable work from $S$, given by $\Delta F_S= F(\rho_S, H_S) - F(\omega_{\beta}(H_S),H_S)$, will in general scale with the size of $n^3$,  as the free energy is an extensive quantity. Hence, in the limit of large $n$ we have that both $\Delta F_S/ \Delta F^{\rm (irr)}$ and $\Delta F_S/ \Delta F^{\rm (res)}$ scale as $O(1/n)$, and hence disappear in the macroscopic limit. 
Thus, the above corrections become negligible in the limit of large systems.  
In other words,
macroscopic phenomenological thermodynamics is insensitive to the strength of
the underlying interactions; making these effects only relevant for small
systems.

\section{Power}
 Although non-zero interactions between $S$ and $B$ tend to
increase dissipation,  they can help to enhance power of an engine, as they can
decrease the time scale of thermalisation $\tau$. Following,  we  provide an
upper bound for the power enhancement of a Carnot-like engine due to the
interaction strength $g$.  In order to do so, we need some considerations on
how $\tau$ is related to $g$. A dimensional analysis argument rapidly suggests
that  $\tau \propto g^{-1}$, and through a more careful analysis implemented in Appendix \ref{Sec:EquilibrationTime} we obtain  $\tau \geq \delta Q /gr$, with $\delta Q$ being
the energy change of $B$ during the equilibration and $r\coloneqq
\norm{[H_B,V]}$ the maximum rate in which $S$ and $B$ can exchange energy.
These considerations allow us to obtain the following bound
\begin{equation}\label{eq:power}
P\coloneqq \frac{W}{\Delta t}\le \frac{g\hspace{0.5mm}r_c\hspace{0.5mm} \eta(g)}{1-\eta +{r_c}/{r_h}}< g\hspace{0.5mm}r_h\hspace{0.5mm} \eta(g) \, ,
\end{equation}
where $W$ is the work produced in a cycle,
$\Delta t$ the cycle time,
$\eta$  the efficiency of the machine,
$r_{c/h}\coloneqq \norm{[H_{B_{c/h}},V_{c/h}]}$ is the maximum rate at which the cold/hot bath with Hamiltonian $H_{B_{c/h}}$ can lose/gain energy and $V_{c/v}$ is the interaction that couples $S$ to the cold/hot bath  (see for details Appendix \ref{Sec:EquilibrationTime}). 
Using  \eqref{eq:efficiencywithg} we can expand \eqref{eq:power} in $g$, obtaining 
\begin{align}
P(g) \le g  r_c\, \eta^{{\rm C}}/(1-\eta^{{\rm C}}+r_c/r_h)-O(g^3).
\end{align}
This result suggests that  the power of an engine  first increases with $g$, reaches a maximum, to then decrease for larger values of $g$ (see also discussion of the limit $g\rightarrow \infty$ in Appendix \ref{Sec:WorkExtraction}). This behaviour is also observed in other treatments of heat engines in strong coupling for various figures of merit \cite{Gelbwaser2015,Uzdin2016,Kkatz2016,Strasberg2016,Newman2016,Xu2016}.

Let us point out that while our bound for $P$ is valid for arbitrary systems, it is expected to be very crude in general, as it depends on the norm of the interaction, $\norm{[H_{B_{c/h}},V_{c/h}]}$. Further progress could be made by either considering particular models or by obtaining better bounds on the time scales of equilibration of generic systems, a notoriously hard and diverse problem \cite{Gogolin2016}. 
Finally, note that relations between power and efficiency of Carnot engines have also been  obtained
in Refs.\ \cite{Karen2013,Naoto2016,Naoto2017}, yielding complementary results.

\begin{figure}
 \includegraphics[width=0.9\columnwidth]{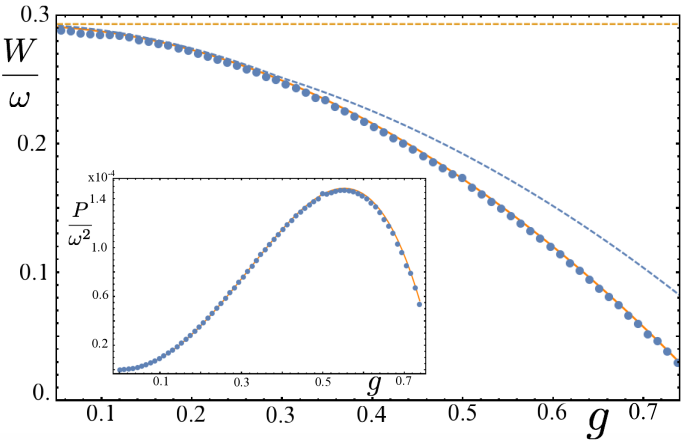}
   \caption{ We take  $\rho^{(0)} = \omega_{\beta_S}(H_S) \otimes \omega_\beta(H_B)$, with $\beta_{S} \neq \beta$, and consider the protocol for maximal work extraction in the weak coupling regime: A quench $\omega\mapsto \omega \beta/\beta_S$,
$m\mapsto \beta_S/\beta$, followed by an isothermal process back to $H_S$. We model the isothermal process by $N=200$ quenches, with a waiting time  $10/g^2$ when computing the  unitary dynamics.  {\it Outer figure:} Work vs.\ strength of interaction.  (i) Blue dots: Extracted work calculated through the unitary evolution of $SB$, (ii) orange line: Same computation but assuming \eqref{eqstrong}, (iii) dashed blue: Optimal protocol by numerically optimising $W$, (iv) dashed orange: $W^{\rm (weak)}$ (i.e., $g=0$).
{\it Inner figure:} Power vs.\ interaction.
Blue dots: exact unitary evolution. Orange line: effective description using our framework.
Parameters for both figures: Setting $\omega=1$, $B$ consists of  $n=165$ oscillators with $m_k=1$ and the $\omega_k$'s uniformly distributed up to $\Omega=1.2$, and $\beta=3.5$. For $S$, $\beta_S =1$, and $m=1$. 
}
\label{PictureWorkExtraction}
\end{figure}

\section{Thermodynamic protocols within the Caldeira-Leggett model}
 We now illustrate our findings by applying them to the model of quantum Brownian motion, captured in the standard Ullersma or
Caldeira-Leggett (CL) model \cite{Ullersma,Caldeira1983}.
In this model, 
\begin{align}
H=H_S+gV+H_B+H_L,
\end{align}
where $S$ is an harmonic oscillator,
$H_S= (m\omega^2 x^2 +{p^2}/{m})/2, $
$B$ a bosonic bath
, $H_B= \sum_{k} (m_k \omega_k^2 x_k^2+p_k^2/ m_k)/2$, which is linearly coupled to $S$ through $gV$ with $V= x\sum_{k} g_k x_k$ and where $g$ quantifies the strength of the interaction, and finally $H_L=x^2 g^2 \sum_{k} {g_k^2}/({m_k \omega_k^2})$
is a renormalization term. Indeed, 
for Ohmic spectral densities, $g$ at the same time quantifies the
deviation from \emph{Markovian dynamics} \cite{OptomechanicalBrownianMotion}.
The couplings $g_k$ are determined by the spectral density $J(w) = ({\pi}/{2} )\sum_k ({g_k^2}/{\omega_k^2}) \delta(\omega-\omega_k),$
which in the continuum limit is often
taken to be well-approximated by an
Ohmic function, $J(\omega)\propto \omega$, for low frequencies until some cut-off $\Omega>0$.
 This model plays
 a crucial role in the study of open quantum systems
\cite{Weiss93}
 and finds numerous applications in thermodynamics \cite{Hanggi2008,EisertPlenioBrownianMotion,Theo2000,Theo2002,Mahler2010,Lutz2011,
 Carrega2015,Philbin2016,Strasberg2016,Newman2016}.

The equilibrium state of $S$ in the CL-model is given by $\Tr_B (\omega_{\beta} (H) )$, irrespectively of $g$, thus satisfying
\eqref{eqstrong}
\cite{Ambegaokar2007,Subasi2012} \footnote{Numerical evidence also shows that the energy contained in the interaction $V$ also equilibrates to the reduced of a global thermal state.}.  
Furthermore, the total Hamiltonian is quadratic, and hence can be solved exactly with matrices of order $O(n^2)$, where $n$ is the number of oscillators in the bath (see, e.g., Refs.\ \cite{Rivas2010,AreaReviewE}). This allows us to numerically simulate thermodynamic protocols exactly for arbitrary strong coupling and large (but finite) baths. Details on the discretisation of the CL model and its simulation are provided in Appendix \ref{Sec:CLmodel}. There, the equilibration time is also discussed, finding $\tau \propto 1/g^2$, for $g\leq 1$, using techniques from Refs.\ \cite{ArnauEquilibration,BerlinEquilibration}, a result which agrees with standard perturbative approaches in the macroscopic limit \cite{Weiss93}.

 We now illustrate our results for  work extraction using the CL model. Crucially,  the first order corrections in \eqref{X^1g} and \eqref{X^Ng} vanish, as the thermal state of $H_B$ is symmetric under  $x_k \leftrightarrow -x_k$. This implies that the optimal protocol in the weak coupling regime is in fact also optimal for small but non-zero $g$. This is perfectly illustrated in Fig.\ \ref{PictureWorkExtraction}, where we plot the work extracted using the weak coupling protocol and the optimal one, which is  obtained by numerically minimising $\Delta F^{(\rm res)/(irr)}$. It is clear that  differences between the two start appearing only at higher orders than $O(g^2)$. 
 Note also that
Fig.\ \ref{PictureWorkExtraction} shows an excellent agreement
between the exact unitary dynamics and our framework, in which
 \eqref{eqstrong} is assumed, even when many quenches are performed.

Now we turn to the question of power.
Here we keep the number of
  quenches $N$ fixed and vary the coupling strength $g$. Since
we deal with isothermal processes, for which $N \rightarrow \infty$,
we take $N$ large but finite.
As a result of the equilibration time $\tau \propto 1/g^2$  for $g\leq 1$, the power
$P(g)=W(g)/\tau(g)$, scales as $P(g) \propto
g^{2} W^{\text{(weak)}}-O(g^{3})$.  This relation is shown in
Fig.\ \ref{PictureWorkExtraction}, where we see that $P(g)$ first
increases as $g^{2}$, 
 reaches a maximum, and then decays to zero for large $g$.

\section{Conclusion}
Bringing together arguments from quantum
thermodynamics and the theory of equilibration in closed many-body
systems, we have derived general strong coupling corrections to the 
second law of thermodynamics. These corrections are applicable to any model of interest, and have been obtained by designing optimal thermodynamic protocols in the strong coupling regime.
The
corrections become relevant if the working body is a small
system, and vanish in first order with the interaction  strength. An upper bound on the power enhancement due to the interaction strength has also been derived.
A particularly relevant open problem is to extend these considerations to scenarios where the system is simultaneously strongly coupled to more than one thermal bath. In this case reaction coordinate mappings \cite{Strasberg2016,Newman2016} appear as a promising technique to extend results in the weak coupling limit \cite{Brunnervirtualtemp}.
It is the hope that this work further stimulates
the emerging field of strong-coupling quantum thermodynamics, aiming at identifying the
potential and burden coming along with such interactions.

\section{Acknowledgements.}
We thank Ignacio Cirac, Andras Molnar, and Tao Shi for insightful discussions. We also thank P. Strasberg, G. Schaller and D. Gelbwaser-Klimovsky for useful comments on the manuscript.
HW acknowledges funding from the ERC
(TAQ)
and the Studienstiftung des Deutschen Volkes, MPL
from the Alexander von Humboldt foundation, RG from the
DFG (GA 2184/2-1). JE thanks for support by the DFG (CRC 183, EI 519/7-1),
the ERC (TAQ), the EU (AQuS), and
the Templeton Foundation.
AR is supported by the Spanish MINECO (QIBEQI FIS2016-80773-P, FISICATEAMO FIS2016-79508-P and Severo Ochoa Grant SEV-2015-0522), Fundaci\'o Privada Cellex, Generalitat de Catalunya (Grant SGR 874, 875, and CERCA Programme), the EC [FP7-ICT-2011-7, Grant No. 288263; OSYRIS (ERC-2013-AdG Grant 339106)] and the Beatriu de Pin\'os fellowship (BP-DGR 2013).
All authors are grateful for support
from the EU COST Action MP1209 on ``Thermodynamics
in the quantum regime''.

\newpage

\appendix
\section*{Appendix}
This Appendix contains eight sections. In Section  \ref{Sec:Notation} and \ref{Sec:Preliminaries},  notation and the basic mathematical tools are introduced. Section  \ref{Sec:WorkExtraction}  contains all results concerning maximal work extraction in the strong coupling regime. Section  \ref{Sec:Heat} deals with heat dissipation, Section  \ref{Sec:CarnotEngines} with heat engines, and Section  \ref{Sec:EquilibrationTime} with power. Finally, in Section  \ref{Sec:CLmodel}   the simulation and equilibration times of the Caldeira-Legget model are discussed.

\section{Notation}
\label{Sec:Notation}

For the sake of clarity, here we recall the notation introduced in the main text:
\begin{itemize}
\item $T,\beta$:  Temperature and inverse temperature, respectively.
\item $H_S$, $H_B$, and $V$: Hamiltonian of $S$, $B$ and the interaction, respectively.
\item $H^{(i)}$: Hamiltonian of $SB$ in the $i$-th step of the protocol. It may contain $V$, so that $H^{(i)}=H_S^{(i)}+V+H_B$.
\item $\omega_{\beta}(H)$: Thermal state, 
\begin{equation}
\omega_{\beta}(H)=\exp(-\beta H)/\tr(\exp(-\beta H)). 
\end{equation}
We will often use the shorthand notation $\omega^{(i)} :=  \omega_{\beta}(H^{(i)})$.
\item $\mathcal{Z}$:  Partition function, $\mathcal{Z}=\tr (e^{-\beta H})$.
\item $\rho_S^{(i)}, \rho^{(i)}_B$:  Reduced states of $S$ and $B$ at the $i$-th step of the protocol, $\rho_{S/B}^{(i)} = \tr_{B/S} \omega^{(i)}$.
\item $\rho_S, \tilde{H}_S, \tilde{H}^{(0)}$: $\rho_S$ initial state of $S$,  $\tilde{H}_S$ is defined through
$\rho_S=\omega_{\beta}(\tilde{H}_S)$,
and  finally $\tilde{H}^{(0)}=\tilde{H}_S+H_B$. 
\item $F(\rho,H)$: Non-equilibrium free energy, $F(\rho,H)= \tr(H\rho)-T S(\rho )$, with $S(\rho)=-\tr(\rho\ln \rho)$.
\item The \emph{operator norm} $\norm{X}$ of an operator $X$ is given by 
	\begin{align}
		\norm{X} = \sup_{\ket{\psi}} \bra{\psi}X\ket{\psi},
	\end{align}
	for normalized state vectors $\ket{\psi}$. In the finite-dimensional spaces considered here, it is simply the largest spectral value of $X$. If $X$ is Hermitian, then we have $\norm{X}=\max_j |\lambda_j|$, where $\lambda_j$ are the eigenvalues of $X$.
\end{itemize} 

\section{Preliminaries}
\label{Sec:Preliminaries}
Let us also introduce  a few technical tools.

\subsection{Differentiation of matrix functions}
Let $A, B$ be two matrices. Then we can differentiate the trace of some matrix function $f(A)$ as 
 \begin{align}
\label{relationDer}
\frac{d}{dg} \tr(f(A+gB)) \bigg|_{g=0}=\tr(B f'(A)),
\end{align} 
where $f'$ is the derivative of $f$. For example, given $H=H_0+gV$, and with $\mathcal{Z}=\tr(e^{-\beta H})$, $\mathcal{Z}_0=\tr(e^{-\beta H_0})$ we obtain 
\begin{align*}
\frac{d}{dg}F(\omega_{\beta}(H),H) \bigg|_{g=0}=\frac{-T}{\mathcal{Z}_0}\frac{d\mathcal{Z}}{dg} \bigg|_{g=0}= \tr(V\omega_{\beta}(H_0)).
\end{align*}

\subsection{Derivative and expansion of exponential operators}
We will also use  the derivative of exponentials of operators
\begin{align}
	\frac{\mathrm{d}}{\mathrm{d} g} \e^{-\beta H(g)} &= -\beta \int_0^1 \e^{-\beta s H(g)}H'(g) \e^{-\beta (1-s)H(g)}\mathrm{d}s.
	\label{DerOp}
\end{align}
From \eqref{DerOp}, we can easily obtain the first order of the  Dyson series (in imaginary time)
\begin{align}
\label{Dyson}
e^{-\beta(H_0 +gV)}= \hspace{-0.5mm}e^{-\beta H_0} \hspace{-0.5mm} \left(\hspace{-0.5mm}\id\hspace{-0.5mm} -\hspace{-0.5mm}\beta g \int_0^{1} \hspace{-1mm}d \tau \hspace{0.5mm} e^{ \beta \tau H_0} V e^{- \tau \beta H_0} \right)\hspace{-1mm}+\hspace{-1mm}O(g^2).
\end{align}
In what follows we will use the short hand notation for the first term of the expansion,
\begin{align}
Y_{H,\beta} := \int_0^{1} \hspace{-1mm}d \tau \hspace{1mm} e^{ \beta \tau H} Y e^{- \tau \beta H}\hspace{-0.5mm}.
\label{timeevol}
\end{align}
This integral can in fact be carried out. For that, we transform the operators of the Hilbert space into vectors, $\ket{Y}=\sum_{i,j} Y_{i,j} \ket{i,j}$ where $Y=\sum_{i,j} Y_{i,j} \ket{i} \bra{j}$, obtaining 
\begin{align}
\ket{Y_{H,\beta}} &= \int_0^{1} d\tau e^{-\tau \beta H\otimes \id}  e^{\tau \beta \id \otimes H^{T}}  \ket{Y}
\nonumber\\
&=   \int_0^{1} d\tau e^{\tau \beta (  \id \otimes H^{T}-Z\otimes \id)}  \ket{Y}
\nonumber\\
&=   \lim_{\epsilon \rightarrow 0} \frac{e^{ \beta (  \id \otimes H^{T}-H\otimes \id)}-\id }{\beta (  \id \otimes H^{T}-H\otimes \id)+i\epsilon}   \ket{Y}
\label{integral}
\end{align}
where $\epsilon>0$ is introduced to ensure that the result is well defined when the denominator has a zero eigenvalue. Of course, the solution is a vector that should be transformed back to the original operator basis.

\subsection{Perturbations of thermal states and generalised covariance}
In the following we will make use of the \emph{generalised covariance} \cite{Kliesch2013}, also known as Kubo-Mori inner product \cite{Mori1956,Kubo1957,Mori1965} or Bogoliubov inner product \cite{Petz1993}. 
For any state $\rho$ and two observables $A,B$ it is defined as 
\begin{align}
	\mr{cov}_\rho(A,B) \coloneqq \int_{0}^1 \tr\left(\rho^{1-\tau}A \rho^{\tau}B\right) - \tr(\rho A)\tr(\rho B). 
\end{align}
For thermal states $\omega_\beta(H)$ we have 
\begin{align}\label{eq:def_covariance}
	\mr{cov}_{\omega_\beta(H)}(A,B) = &\tr\left(\omega_\beta(H) A_{H,\beta} B\right) \nonumber\\
	&- \tr(\omega_\beta(H)A)\tr(\omega_\beta(H)B).
\end{align}
From a physical point of view, the generalised covariance measures how thermal
perturbation values respond to a change of the Hamiltonian in a thermal state \cite{Mori1956,Kubo1957,Mori1965}.
Let $H_t$ be a smooth family of Hamiltonians. Then it follows from \eqref{DerOp} that
\begin{align}\label{thermal_perturbation}
	\tr\left(A \left.\frac{\mr d}{\mr d t}\right|_{t=0} \omega_\beta(H_t)\right) = -\beta \mr{cov}_{\omega_\beta(H_0)}\left(A,H'_0\right),
\end{align}
where $H'_0 = \mr{d} H_t / \mr{d}t |_{t=0}$. 
 For finite perturbations of the Hamiltonian $H_0$ one can integrate the above equation (see~\cite{Kliesch2013}).

\subsection{Symmetry of relative entropy}
In general, the quantum relative entropy is an asymmetric function, i.e.,
$S(\rho\|\sigma)\neq S(\sigma\|\rho)$. Nevertheless, it is known that it is symmetric up to second order in
the difference between $\rho$ and $\sigma$ (see, e.g., Ref.\  \cite{amari2007methods}). 
The following Lemma, which might be of
independent interest, shows this result when $\rho$ and $\sigma$ are thermal states, and their distance is modified by changing their Hamiltonians. This result will be key for the derivation
of the exact form of the corrections to work and efficiency for small but
finite coupling strengths.
\begin{lemma}[Perturbative symmetry of relative entropy]\label{lemma:relent_symmetry}
Let $H_t$ be a one-parameter family of operators. Then for small $t$, we have
\begin{align}
	\Delta_t \coloneqq S(\omega_\beta(H_0) \| \omega_\beta(H_t)) - S(\omega_\beta(H_t) \| \omega_\beta(H_0)) = O(t^3).
\end{align}
\begin{proof}
In the following we will use $f'_t$ as shorthand for
the derivative of the function $f_t$ with respect of
$t$.  Let us also define the partition function $\mathcal{Z}_t
\coloneqq \tr(\e^{-\beta H_t})$. Using \eqref{DerOp},
we obtain $\log(\mathcal{Z}_t)' = -\beta\tr(\omega_t H'_t)$. We can
then calculate the derivative of $S(\omega_0 \| \omega_t)$ as
\begin{align}\label{eq:derivative_Firr}
	S(\omega_0 \| \omega_t)' &= - \tr\left(\omega_0 \log(\omega_t)'\right) = \beta \tr\left(\omega_0 H_t'\right) + \log(\mathcal{Z}_t)'\nonumber\\ &= \beta\tr\left((\omega_0 -\omega_t) H_t'\right). 
\end{align}
To compute the derivative of $S(\omega_t \| \omega_0)$, let us first compute the derivative of the entropy of $\omega_t$, to get
\begin{align}
S(\omega_t)' &= -\tr\left(\omega_t \log(\omega_t)'\right) - \tr\left(\omega_t' \log(\omega_t)\right) \nonumber\\
	 	&= \beta\tr\left(\omega_t H_t'\right) + \log(\mathcal{Z}_t)' - \tr\left(\omega_t' \log(\omega_t)'\right)\nonumber\\
		 &= -\tr\left(\omega_t' \log(\omega_t)\right).
\end{align}
We then obtain for the derivative of the relative entropy
\begin{align}
S(\omega_t \| \omega_0)' &= - S(\omega_t)' - \tr\left(\omega_t \log(\omega_0)\right)' \nonumber\\
	   &= - \tr\left(\omega_t' \left(\log(\omega_0)-\log(\omega_t)\right)\right) \nonumber\\
	  \label{eq:first_derivative} &=  \beta\tr\left(\omega_t'(H_0 - H_t)\right),
\end{align}
where \eqref{eq:first_derivative} follows from $\Tr(\omega'_t)=0$. Then, we have for the first derivative of $\Delta_t$
\begin{align}
\Delta_t' &= \beta\left[\tr\left((\omega_0 - \omega_t) H_t'\right) - \tr\left(\omega_t'(H_0-H_t)\right) \right].
\end{align}
From this expression we can easily compute the second derivative as
\begin{align}
\Delta_t'' &= \beta\left[\tr \left((\omega_0-\omega_t)' H_t'\right) + \tr\left((\omega_0-\omega_t)H_t'\right)\right] \nonumber\\
        & - \beta\left[\tr\left(\omega_t''(H_0-H_t)\right)+\tr\left(\omega_t'(H_0-H_t)'\right)\right]\nonumber \\
 	 &= \beta\left[\tr\left((\omega_0-\omega_t)H_t'\right) - \tr\left(\omega_t''(H_0-H_t)\right)\right].
\end{align}
In particular, we get $\Delta_0'=0$ and $\Delta_0''= 0$, which proves the claim. 
\end{proof}
\end{lemma}

\section{Maximizing work extraction in the strong coupling regime}
\label{Sec:WorkExtraction}

In this section, we optimize work extraction protocols over all cyclic protocols that can be constructed with the operations (A)-(C) described in the main text, and given the initial non-interacting Hamiltonian $H^{(0)} = H_S + H_B$ and the initial state $\rho^{(0)}=\rho_S \otimes \omega_\beta(H_B)$. Here cyclicity is understood in terms of the Hamiltonian, so that at the end of the protocol consisting of $N+1$ steps we have that $H^{(N+1)} =H^{(0)}$. Let us stress that these $N+1$ steps can in principle consist of any of the operations (A), (B) and (C) of the main text, and $N$ can be arbitrarily large.

We now focus on protocols where the interaction is turned on and off only once in the protocol (i.e., operation (A) is only implemented twice). Later we will show that this is in fact optimal. Any protocol can then always be described as,
\begin{enumerate}
\item A series of quenches are applied to the local Hamiltonian of $S$, $H_S \rightarrow ... \rightarrow H_S^{(1)}$, and the interaction between $S$ and $B$ is turned on. The total Hamiltonian becomes $H^{(1)}=H_S^{(1)}+V+H_B$. The expected work gain of this process reads
\begin{align}
W_{1}=\Tr \left((H_S-H_S^{(1)}-V)\rho^{(0)}\right).
\end{align}

\item  The Hamiltonian of $S$ is modified while $S$ is in contact with $B$, amounting to a sequence of quenches
$H_S^{(1)}\mapsto H_S^{(2)}\mapsto \dots \mapsto H_S^{(N)}$, each followed by an equilibration to the corresponding thermal state, until $H^{(N)}=H_S^{(N)}+V+H_B$ is reached. Note that we  assume that $H_S^{(N)}$ is independent of $N$, so that increasing $N$ means doing the same Hamiltonian transformation in more steps (i.e., slower). During this process the state of 
$S$ reads
$\rho_S^{(j)}=\tr_B(\omega^{(j)})$, $j=1,\dots,N$. The  total work gain is then
\begin{align}
W_2=&\sum_{i=1}^{N} \Tr (\rho_{S}^{(i)}(H_{S}^{(i)}-H_{S}^{(i+1)} ))
\nonumber\\
=&\sum_{i=1}^{N} \Tr (\omega^{(i)}(H^{(i)}-H^{(i+1)} ))
\nonumber\\
=&F(\omega^{(1)},H^{(1)})-F(\omega^{(N)},H^{(N)})
\nonumber\\
&-  T\sum_{i=1}^{N} {\rm S}(\omega^{(i)}\|\omega^{(i+1)}).
\label{isothermal_strong}
\end{align}

\item The interaction between $S$ and $B$ is turned off, and the Hamiltonian of $S$ is brought back to the initial form by an arbitrary series of quenches, $H^{(N)}_S \rightarrow ... \rightarrow H_S$. The expected work gain is simply 
\begin{align}
W_{3}=\Tr \left((V+H_S^{(N)}-H_S)\omega^{(N)}\right).
\end{align}
\end{enumerate}

These three steps conclude a cyclic Hamiltonian process.
Two remarks are now in order. Firstly,  the last term of \eqref{isothermal_strong} is positive and tends to zero in the limit $N\rightarrow \infty$, i.e., for isothermal processes. That implies 
\begin{align}
W_2 \leq W^{\mathrm{(isoth)}}_{\rm sc}=F(\omega^{(1)},H^{(1)})-F(\omega^{(N)},H^{(N)})
\label{Wisostrong}
\end{align}
where $W^{\mathrm{(isoth)}}_{\rm sc}$ stands for the work gain of an isothermal transformation in the strong coupling regime \cite{Gallego14}.  Hence, in the optimal protocol contacts between $S$ and $B$ must be in form of isothermal transformations.

The second remark concerns points 1 and 3. If the interaction is weak, the
energy of turning on and off the interaction can be neglected. Exactly in this case, one can obtain the usual expression of optimal work given by the free energy. This is indeed obtained  with the choice of $H_S^{(N)}=H_S$ and $H_S^{(1)}$ such that $\omega_{\beta}(H_S^{(1)})=\rho_S$. Then one obtains
\begin{align}
W^{\rm (weak)}&= F(\rho_S,H_S)-F(\omega_\beta(H_S),H_S)
\nonumber\\
&=F(\rho^{(0)},H^{(0)})-F(\omega^{(0)},H^{(0)}),
\end{align}
with $\omega^{(0)} := \omega_\beta(H^{(0)})$. However, if the interaction is non-negligible,
we now show that steps 1 and 3 become a source
of irreversibility. As a consequence, the optimal protocol must consist of only one thermal contact between $S$ and $B$ -- hence justifying the form of the protocol considered -- further contacts can only decrease the extractable work.

Now we perform some simple algebra to express the total work of the optimal protocol in a convenient form. Adding up the three work contributions, and adding and subtracting $F(\omega^{(0)},H^{(0)})$, we obtain
\begin{align}
W =& W_{1}+W_{\rm sc}^{(\rm isoth)}+W_{3}
\nonumber\\
=&W^{\rm (weak)}+F(\omega^{(0)},H^{(0)})-F(\rho^{(0)},H^{(0)})\nonumber\\
&+\Tr \left((H_S-H_S^{(1)}-V)\rho^{(0)}\right)+F\left(\omega^{(1)},H^{(1)}\right)
\nonumber\\
&-F\left(\omega^{(N)},H^{(N)}\right)
+\Tr \left((V+H_S^{(N)}-H_S)\omega^{(N)}\right).\nonumber
\end{align}
Using that
$F(\rho^{(0)},H^{(1)})+\Tr  ((H_S-H_S^{(1)}-V)\rho^{(0)} )=F(\rho^{(0)},H^{(0)})$
and $F(\omega^{(N)},H^{(N)})-F(\omega^{(N)},H^{(0)}
)= \Tr ((V+H_S^{(N)}-H_S)\omega^{(N)} ) $ we further obtain
\begin{align}
W&= W^{\rm (weak)}+F(\omega_0,H^{(0)})-F(\rho^{(0)},H^{(1)}) 
\nonumber\\
&\quad +F(\omega^{(1)},H^{(1)})-F(\omega^{(N)},H^{(0)})
\nonumber\\
&= W^{\rm (weak)}- \Delta F^{\rm (res)}  -\Delta F^{\rm (irr)},
\end{align}
where we have defined
\begin{align}
 &\Delta F^{\rm (res)} :=  F(\omega_\beta(H^{(N)}),H^{(0)})-F(\omega_\beta(H^{(0)}),H^{(0)}) ,
\nonumber\\
 &\Delta F^{\rm (irr)} :=  F(\rho^{(0)},H^{(1)})-F(\omega_\beta(H^{(1)}),H^{(1)}) .
  \label{Deltafappend}
\end{align}
By noting that $F(\rho,H)-F(\omega_\beta(H),H)=T {\rm S}(\rho \|\omega_\beta(H))$, we already obtain that always 
$W \leq W^{\rm (weak)}$, so that interactions are detrimental for work extraction.
However, at the moment, the terms \eqref{Deltafappend}  depend on $H_S^{(1)}$ and $H_S^{(N)}$, i.e., on the particular points where $S$ is connected and disconnected from $B$.
In order to maximise $W$, we hence need to minimize $\Delta F^{\rm (res)}$ and $\Delta F^{\rm (irr)}$ over $H_S^{(1)}$ and $H_S^{(N)}$, i.e., 
\begin{align}
&W_{\rm max} = W^{\rm (weak)} - \Delta F^{\rm (res)}_{\rm min} -\Delta F^{\rm (irr)}_{\rm min}   ,
\label{minimisationsDeltaF}
\end{align}
with
\begin{align}
\label{mininitial}
& \Delta F^{\rm (irr)}_{\rm min} =\min_{H_{S}^{(1)}} \Delta F^{\rm (irr)}, \nonumber\\
&\Delta F^{\rm (res)}_{\rm min}= \min_{H_{S}^{(N)}} \Delta F^{\rm (res)}. 
\end{align}
We now proceed to solve these minimizations, which can be carried out independently.
For that, we can use that in a (local)
minimum of a function, its derivative must vanish. Since here we minimise over a matrix,  the relation \eqref{relationDer} becomes useful.

\subsection{Minimization of $\Delta F^{\rm (irr)}$}
Let $X_S$ be the choice of $H_S^{(1)}$ yielding $\Delta F^{\rm (irr)}_{\rm min}$,  and define 
\begin{equation}
X(t)\coloneqq X_S\otimes \id_B+t Y_S\otimes \id_B+V+\id_S \otimes H_B\, .
\end{equation}
Then, for any $Y_S$, it must hold that 
\begin{equation}
\label{derivat}
\frac{d}{dt} \left( F\left(\rho^{(0)},X(t)\right)-F\left(\omega_\beta(X(t)),X(t)\right)  \right) \bigg|_{t=0}=0 . \hspace{5mm} 
\end{equation}
That is, if we perturb the solution $X_S$ by $t Y_S$, then the derivative w.r.t $t$ must be zero for all $Y_S$. In other words, we are standing in a minimum. Conversely, we can use  \eqref{derivat} to find $X_S$.
Indeed, computing \eqref{derivat} using \eqref{relationDer},  we obtain
\begin{align}
\frac{d}{dt}  &\left[F(\rho^{(0)},X(t))-F(\omega_\beta(X(t)),X(t))  \right] \bigg|_{t=0}
\nonumber\\
&\quad=\tr \left(\left(\rho^{(0)}-\omega_{\beta}(X(0))    \right) Y_S  \otimes \id_B\right) 
\nonumber\\
&\quad= \tr_S \left( \left( \rho_S- \tr_B \left[ \omega_{\beta}(X(0))\right] \right)  Y_S \right) =0.
\end{align}
Since this holds for any  $Y_S$, and setting $X:=X(0)$, it follows that
\begin{align}
\rho_S = \tr_B \left(\omega_{\beta}(X)  \right) .
\label{solX_S}
\end{align}
This matrix equation (of the size of $S$) implicitly provides $X_S$, and hence also $\Delta F^{\rm (irr)}_{\rm min}$.

We now show that there is at least one minimum of $\Delta F^{\rm (irr)}$, so that
\eqref{solX_S} always provides the desired solution, provided that $\rho_S$ is
a full rank state. 
For that, we first show  that for any full rank state
	$\rho_S$ and non-trivial Hamiltonian $H_S^{(1)}$, the term $\Delta F^{\rm (irr)}$ diverges with the operator norm  $H_S^{(1)}$.  
More explicitly, let us parametrize the Hamiltonians as $H_S^{(1)}= \lambda \hat H_S^{(1)}$ with $\hat H_S^{(1)}$ controlling its direction, i.e.
$\| \hat H_S^{(1)}\|=1$, and $\lambda$  its operator norm.
Let $P_S^{(1)}$ denote the ground-state space of $\hat H_S^{(1)}$. Then the state
\begin{equation}
	\lim_{\lambda\to\infty} \omega_\beta(\lambda \hat{H}_S^{(1)}+V+H_B) \coloneqq \sigma_{SB}
\end{equation}
is supported only with the subspace $P_S^{(1)} \otimes \mc H_B$, where $\mc H_B$ denotes the Hilbert-space of the bath. 
But then we have
\begin{align}
	\lim_{\lambda\to\infty} \Delta F^{\mathrm{(irr)}} = \frac{1}{\beta} S\left(\rho^{(0)}\otimes \omega_\beta(H_B) \| \sigma_{SB}\right)= \infty,
\end{align}
since the relative entropy diverges if the support of the first argument is not contained in that of the second. 
Now, the function $\Delta F^{\mathrm{(irr)}}$ is (i) positive, (ii) continuous and
differentiable, (iii) and tends to $+\infty$ for large Hamiltonians
$H^{(1)}_S$. It follows from (i)-(iii) that $\Delta F^{\rm (irr)}$ has at least
one minimum, which can be obtained through \eqref{solX_S}. In case that the
solution of \eqref{solX_S} is not unique and there is more than a minimum, the
global one is chosen.

\subsection{Minimization of $\Delta F^{\rm (res)}$}

Let us now proceed similarly for $\Delta F^{\rm (res)}_{\rm min}$, a calculation which turns out to be bit more involved. 
Similarly as in the last section, assume that $Z_S$ is the solution to $\min_{H_{S}^{(N)}} \Delta F^{\rm (res)}$, and $tY_S$ is some perturbation over this solution.  Then, we define
\begin{equation}
Z(t)=Z_S \otimes \id_B +t Y_S \otimes \id +V +\id_S \otimes H_B,
\end{equation}
and $\mathcal{Z}(t)=\tr \left(e^{-\beta Z(t)} \right)$ and $\omega_\beta (t)\coloneqq \omega_\beta(Z(t))$. 
The condition of minimum implies
\begin{align}
\frac{\dd}{\dd t} \Delta F^{\rm (res)} &= \frac{\dd}{\dd t} \bigg|_{t=0}\frac{1}{\beta}S\left(\omega_\beta(t) \| \omega_\beta(H^{(0)})\right) \nonumber\\ 
			 &= 0 \quad \forall \ \ Y_S\, . \nonumber
\end{align}
From the calculation in the proof of Lemma~\ref{lemma:relent_symmetry}, we find that
\begin{align}
	\frac{\dd}{\dd t} \Delta F^{\rm (res)} &= \beta\tr\left((H^{(0)} - Z(0))\left.\frac{\dd \omega_\beta(t)}{\dd t}\right|_{t=0}\right). 
\end{align}
We can now use the perturbation-formula \eqref{thermal_perturbation} and obtain
\begin{align}
	\frac{\dd}{\dd t} \Delta F^{\rm (res)} &= \mathrm{cov}_{\omega_\beta(0)}\left(H^{(0)}-Z(0),Y_S\right) \overset{!}{=} 0.
\end{align}
Setting $Z\coloneqq Z(0)$ and writing out the definition of the generalised covariance, we then obtain
\begin{align}
	&\tr(\omega_\beta(Z) Y_S)\tr\left((H^{(0)}_{Z,\beta} - Z) \omega_\beta(Z)\right)\nonumber\\
		&\quad\quad\quad\quad= \tr\left(\omega_\beta(Z) (H^{(0)}_{Z,\beta}-Z) Y_S\right).
\end{align}
Since this condition has to hold for all $Y_S$, it is, per definition of the partial trace, equivalent to 
\begin{align}
	&\tr_B(\omega_\beta(Z))\tr\left((H^{(0)}_{Z,\beta} - Z) \omega_\beta(Z)\right)\nonumber\\
	&\quad\quad\quad\quad= \tr_B\left(\omega_\beta(Z) (H^{(0)}_{Z,\beta}-Z)\right).\label{conditionZS}
\end{align}
This is the desired implicit solution for $Z_S$.

To finish this section, we show that the residual free energy $\Delta F^{({\rm res})}$ has at least one local minimum, and hence \eqref{conditionZS} always provides a solution. Unlike $\Delta F^{{\rm (irr)}}$, 
 $\Delta F^{({\rm res})}$ does not diverge to $+\infty$ with the operator norm of $H_S^{(N)}$, but tends to a constant.
As above, the Hamiltonian is parametrised as $H_S^{(N)}= \lambda  \hat H_S^{(N)}$, where
$\lambda$ controls its norm and $\hat H_S^{(N)}$ controls its direction, i.~e.
$\| \hat H_S^{(N)}\|=1$. As in the last section, let $P_S^{(N)}$ denote the ground-state subspace of $\hat H_S^{(N)}$. 
In the limit $\lambda \to \infty$, 
the thermal state $\omega_\beta^{(N)}(\lambda)=\omega_\beta(\lambda\hat H^{(N)}_S+V+H_B)$ tends to the state
\begin{equation}
	\lim_{\lambda \to \infty} \omega_\beta^{(N)}(\lambda) = \sigma_{SB}^{(N)},
\end{equation}
which is again supported within the subspace $P_S^{(N)}\otimes \mc H_B$.
Since $\Delta F^{\rm (res)} = F_\beta(\omega_\beta^{(N)}(\lambda),H^{(0)})-F_\beta(\omega_\beta(H^{(0)}),H^{(0)})$, we then have
\begin{align}
	\lim_{\lambda\to \infty} \Delta F^{\rm (res)}  &=F_\beta(\sigma_{SB}^{(N)},H^{(0)}) - F_\beta(\omega_\beta(H^{(0)}),H^{(0)})\nonumber\\
						&\leq \norm{H^{(0)}} - F_\beta(\omega_\beta(H^{(0)}),H^{(0)}).\nonumber
\end{align}
In order to see that the minimum of the residual free energy
takes place for a finite $\lambda$ we show that the asymptotic value above is approached from
below, and equivalently that the derivative for large $\lambda$ tends to 0 from the positive side.
The first derivative of the residual free energy reads
\begin{equation}
\frac{\dd F(\omega^{(N)}(\lambda),H^{(0)})}{\dd \lambda}=
\tr\left((H^{(0)}-H^{(N)}(\lambda))\frac{\dd \omega^{(N)}(\lambda)}{\dd \lambda} \right),
\end{equation}
where $\omega^{(N)}(\lambda)\coloneqq \omega_\beta(\lambda \hat H^{(N)}_S + V + H_B)$. 
We can use \eqref{thermal_perturbation} to obtain
\begin{align}
\frac{\dd F(\omega^{(N)}(\lambda),H^{(0)})}{\dd \lambda} &= -\beta \mathrm{cov}_{\omega_\beta(\lambda)} \left(H^{(0)}- H^{(N)},\hat H^{(N)}_S\right)\nonumber\\
&= \lambda \beta \mathrm{cov}_{\omega_\beta(\lambda)}\left(\hat H^{(N)}_S,H^{(N)}_S\right) - \nonumber \\
&\ \ \beta   \mathrm{cov}_{\omega_\beta(\lambda)} \left(H^{(0)}_S-V,\hat H^{(N)}_S\right).\label{eq:minimumres}
\end{align}
In the limit $\lambda\to\infty$, we have
\begin{align}
	\lim_{\lambda\to\infty}\mathrm{cov}_{\omega_\beta(\lambda)}&\left(\hat H^{(N)}_S,H^{(N)}_S\right)\nonumber\\
	&= \mathrm{cov}_{\sigma_{SB}^{(N)}}\left(\hat H^{(N)}_S,H^{(N)}_S\right)\\
	&=0,
\end{align}
since $\sigma_{SB}^{(N)}$ is supported within the ground-state space of $H^{(N)}_S$. 
Furthermore, in the limit $\lambda\rightarrow \infty$, we have $\exp(-\tau\lambda \hat H^{(N)}_S+V+H_B)\hat H^{(N)}_S =0$ for any $\tau>0$.
This then also implies
\begin{align}
	\lim_{\lambda\to\infty} \mathrm{cov}_{\omega_\beta(\lambda)} \left(H^{(0)}_S-V,\hat H^{(N)}_S\right)=0.
\end{align}
We can then expand an expression of the form 
\begin{align}
\mathrm{cov}_{\omega_\beta(\lambda)} \left(A,\hat H^{(N)}_S\right)
\end{align}
in $y=\exp(-\beta\lambda)$ around $y=0$. 
The above considerations show that the first order vanishes, hence for very large $\lambda$ we have
\begin{align}
	\mathrm{cov}_{\omega_\beta(\lambda)} \left(A,\hat H^{(N)}_S\right) = \e^{-\beta \lambda} f(A) + O(\e^{-2 \beta \lambda}).
\end{align}
Since 
\begin{align}
\mathrm{cov}_{\omega_\beta(\lambda)} \left(\hat H^{(N)}_S ,\hat H^{(N)}_S \right)\geq 0
\end{align}
for all $\lambda$,  we have $f(\hat H^{(N)}_S)\geq 0$. 
We can now use these results in Eq.~\ref{eq:minimumres} and find for large $\lambda$ 
\begin{align}
	\frac{\dd F(\omega^{(N)}(\lambda),H^{(0)})}{\dd \lambda} = \e^{-\beta\lambda} \left(\lambda f(\hat H^{(N)}_S) - f(H^{(0)}-V)\right),\nonumber 
\end{align}
up to terms $O(\e^{-2\beta\lambda})$ and higher. 
We hence see that the derivative of $\Delta F^{\rm (res)}$ becomes positive for large enough $\lambda$ and  $\Delta F^{\rm (res)}$ approaches its limiting value from below for large $\lambda$. Since it is a positive, smooth function, this implies the existence of at least one minimum  and \eqref{conditionZS} always provides a solution to the problem
of maximising work extraction. Again, if several solutions exist, the global optimum must be chosen. 

In sum, so far we have reduced the problem of finding the optimal protocol
for work extraction to the solution of two matrix equations, \eqref{solX_S} and
\eqref{conditionZS}. We now provide the solution of these
equations, and hence an expression for  $W_{\rm max}$ in
\eqref{minimisationsDeltaF}, at lowest non-vanishing order in the interaction
strength. For that, let us replace $V$ by $gV$, and expand  relevant quantities over $g$. 

\subsection{Expansion in orders of $g$}

Recall that $X_S$ and $Z_S$ are the local Hamiltonians on $S$ minimizing $\Delta F^{(\rm irr)}$ and $\Delta F^{(\rm res)}$ respectively, and also
\begin{align}
\label{eq:def_local_solutionX}X&:=X_S  + gV+ H_B, \\
\label{eq:def_local_solutionZ}Z&:=Z_S +gV+ H_B.
\end{align}
The Hamiltonians $X_S$ and $Z_S$ are functions of $g$ since they
depend on the interaction given by $gV$. This can be expressed by defining the
functions $X_S(g)$ and $Z_S(g)$, which we can formally expand in powers of
$g$ as 
\begin{align}
X_S&=X_S(0)+ gX'_S(0) + O(g^2),\\
Z_S&=Z_S(0)+ gZ'_S(0) + O(g^2),
\end{align}
where $X'_S$ is the derivative with respect to $g$ of $X_S(g)$ and equivalently for $Z_S(g)$. Using \eqref{eq:def_local_solutionX} and \eqref{eq:def_local_solutionZ} we can write an expansion in $g$ for the Hamiltonians 
\begin{align}
\nonumber X&=X_S(0)+ H_B + g(V+X'_S(0))+ O(g^2)\\
\nonumber &:= X_0 +gX_1+O(g^2),\\
\nonumber Z&=Z_S(0)+H_B + g(V+Z'_S(0))+ O(g^2)\\
\label{eq:def_expansion_YZ}&:= Z_0 +gZ_1+O(g^2).
\end{align}
Consider also the expansions of the corresponding thermal states,
\begin{align}
\nonumber &\omega_{\beta}(X)= \omega_0^{(X)} + g \omega_1^{(X)} + g^2 \omega_2^{(X)}+O(g^3),\\
&\omega_{\beta}(Z)= \omega_0^{(Z)} + g \omega_1^{(Z)} + g^2 \omega_2^{(Z)}+O(g^3).
\label{ExpExpXZ}
\end{align}
Note that, from \eqref{solX_S} and \eqref{conditionZS}, it is clear that $Z_0=H^{(0)}$, $X_0 = \tilde{H}^{(0)}$, and hence $\omega_0^{(Z)}=\omega_{\beta}(H^{(0)})$, $\omega_0^{(X)}=\omega_{\beta}(\tilde{H}^{(0)})$. Moreover, we have that,
\begin{align}
\Delta F^{\rm (irr)}_{\rm min} &=  T  S\left(\omega_{\beta}(\tilde{H}^{(0)}) || \omega_{\beta}(X)\right) ,
\\
\Delta F^{\rm (res)}_{\rm min} &= T  S\left(\omega_{\beta}(Z)||\omega_{\beta}(H^{(0)}) \right).
\end{align}
Let us now expand these expressions.  The calculation follows from the proof of Lemma~\ref{lemma:relent_symmetry} and yields
\begin{align}
\frac{d \Delta F^{\rm (irr)}_{\rm min}}{dg}  = -g \tr\left(X_1 \omega_1^{(X)} \right) +O(g^2)
\end{align}
and
\begin{align}
\frac{d^2 \Delta F^{\rm (irr)}_{\rm min}}{dg^2} \bigg|_{g=0} = - \tr\left(X_1 \omega_1^{(X)} \right).
\label{secondderIrr}
\end{align}
Proceeding analogously for $ \Delta F^{\rm (res)}$, we obtain
\begin{align}
&\frac{d \Delta F^{\rm (res)}_{\rm min}}{dg} =  -g \tr\left(Z_1 \omega_1^{(Z)} \right) +O(g^2),
\nonumber\\
&\frac{d^2 \Delta F^{\rm (res)}_{\rm min}}{dg^2} \bigg|_{g=0}=- \tr\left(Z_1 \omega_1^{(Z)} \right).
\label{secondderRes}
\end{align}
Importantly, from these calculations we learn that (i) the first non-vanishing  terms are of order $O(g^2)$ and only depend on first order corrections of $Z$ and $X$, and that (ii) the two correcting terms have the same behavior at second order in $g$, which is a consequence of Lemma~\ref{lemma:relent_symmetry}.

We now compute explicitly the expansions \eqref{eq:def_expansion_YZ} and \eqref{ExpExpXZ}. We start by computing the expansion of $X$ as determined by the solution \eqref{solX_S}, which, expanded using \eqref{eq:def_expansion_YZ} and \eqref{ExpExpXZ}, can be simply written as
\begin{equation}
\label{eq:condition_X_first_order}0=\tr_{B}(\omega_1^{(X)}):=\tr_B \left(\frac{d \omega_{\beta}(X(g))}{dg} \bigg|_{g=0}\right).
\end{equation}
From \eqref{DerOp} and \eqref{timeevol}, and using the cyclic property of the trace, we obtain
\begin{align}
T \frac{d \omega_{\beta}(X(g))}{dg} \bigg|_{g=0}\hspace{-1mm}=\omega_0^{(X)} \tr\left(X_1 \omega_0^{(X)}\right)-\left(X_1 \omega_0^{(X)}\right)_{X_0}. \nonumber
\end{align}
Taking the partial trace over $B$, one can express \eqref{eq:condition_X_first_order}  as
\begin{align}
\tr (\omega_0^{(X)} X_1) \one_S =\tr_{B} (X_1 \omega_\beta (H_B))_{X_S(0)}.
\end{align}
Recalling that $X_1=X'_S(0) +V$ and defining $V':=-\tr_B(V \omega_{\beta}(H_B))$ one can re-express the equation above as
\begin{align}
L(X'_S(0))=L(V'),
\end{align}
where $L$ is a linear super-operator defined as $L(\cdot) := \tr(\omega_{\beta}(\tilde{H}_S) \:\:\cdot\:\:) \one - (\:\cdot\:) _{\tilde{H}_S}$. The solution is then unique, given by $X_S'(0)=V'=-\tr_B(V \omega_{\beta}(H_B))$ up to addition of any operator $M$ fulfilling $L(M)=0$. We will later see that adding any operator $M$ does not alter the correction to work, hence all the possible solutions perform equally good regarding work extraction. Altogether we find that
\begin{align}
\nonumber X_S&=\tilde{H}_S - g(\tr_B(V \omega_{\beta}(H_B)) -M) +O(g^2),\\
\label{eq:finalsolutionX}X_1&=V - \tr_B (V\omega_{\beta}(H_B))+ M +O(g^2),
\end{align}
for any $M$ fulfilling $L(M)=0$. In practice, since $M$ does not contribute to the extracted work, we can just set $M=0$.

We can proceed analogously for $\Delta F^{\rm (res)}$, expanding \eqref{conditionZS} at first order in $g$. One obtains an equivalent condition as for the case of $X_S$, that is, one obtains 
\begin{align}
\tr (\omega_0^{(Z)} Z_1) \one_S =\tr_{B} (Z_1 \omega_\beta (H_B))_{Z_S(0)},
\end{align}
which can be again re-expressed as
\begin{equation}
R(Z_S'(0))=R(V'),
\end{equation}
where $R(\cdot) := \tr(\omega_{\beta}(H_S^{(0)}) \:\:\cdot\:\:) \one - (\:\cdot\:) _{H^{(0)}_S}$. This in turn yields the solution 
\begin{align}
\nonumber Z_S&=H_S^{(0)} - g(\tr_B(V \omega_{\beta}(H_B)) -M') +O(g^2),\\
\label{eq:finalsolutionZ} Z_1&=V - \tr_B (V\omega_{\beta}(H_B))+ M' +O(g^2),
\end{align}
for any $M'$ fulfilling $R(M')=0$. 

We now use the solutions for the optimal Hamiltonians $X_1$ and $Z_1$ at first order, given by \eqref{eq:finalsolutionX} and \eqref{eq:finalsolutionZ}, to obtain the lowest order corrections to $\Delta F^{\rm (irr)}_{\rm min}$. Introducing the dressed interaction 
\begin{align}
\tilde{V}:=V-\Tr_B\left(V \omega_{\beta}(H_B)\right)
\end{align}
we obtain families of solutions given by
\begin{align}
	X_1 = \tilde{V}+M,\quad Z_1 = \tilde{V}+M',
\end{align}
with any $M$ and $M'$ fulfilling $L(M)=0$ and $R(M')=0$. Using the perturbation formula \eqref{thermal_perturbation}, we then have
\begin{align}
\nonumber\tr\left(X_1 \omega_1^{(X)}\right) &= -\beta\mathrm{cov}_{\omega_\beta(\tilde{H}^{(0)})}(\tilde{V}+M,\tilde{V}+M)\\
\label{eq:independentM}&=-\beta\mathrm{cov}_{\omega_\beta(\tilde{H}^{(0)})}(\tilde{V},\tilde{V}),\\
\nonumber\tr\left(Z_1 \omega_1^{(Z)}\right) &= -\beta\mathrm{cov}_{\omega_\beta(H^{(0)})}(\tilde{V}+M',\tilde{V}+M')\\
\label{eq:independentM'}&= -\beta\mathrm{cov}_{\omega_\beta(H^{(0)})}(\tilde{V},\tilde{V}).
\end{align}
Eq.\ \eqref{eq:independentM} follows since $L(M)=0$ implies that $\tr(\omega_\beta (\tilde{H}^{(0)}) M) \one = M_{\tilde{H}^{(0)}}$ and from the the definition of the generalised covariance \eqref{eq:def_covariance}. An equivalent argument can be used to show \eqref{eq:independentM'}.
Finally, using \eqref{secondderIrr} and \eqref{secondderRes}, we reach our main result
\begin{align}
\Delta F^{\rm (irr)}_{\rm min}= \frac{\beta g^2}{2} \mathrm{cov}_{\omega_\beta(\tilde{H}^{(0)})}(\tilde{V},\tilde{V}) +O(g^3),
\label{irrweakcouplingsolution}
\\
\Delta F^{\rm (res)}_{\rm min}= \frac{\beta g^2}{2}  \mathrm{cov}_{\omega_\beta(H^{(0)})}(\tilde{V},\tilde{V})+O(g^3).
\label{weakcouplingsolution}
\end{align}

Let us now summarize our results. The problem we have addressed is to maximise work extraction in a cyclic Hamiltonian process under the set of operations $(A)-(C)$ of the main text, and for an initial non-interacting Hamiltonian $H^{(0)} = H_S + H_B$ and the initial state $\rho^{(0)}=\rho_S \otimes \omega_\beta(H_B)$. Firstly, by expressing the maximal work as the difference between the maximal extractable work in the weak coupling regime and two positive terms \eqref{minimisationsDeltaF}, we have shown that strong interactions can only be detrimental for work extraction. More importantly, we have reduced the initial optimization over all protocols to two minimizations over local Hamiltonians as given by \eqref{mininitial}. These two minimizations have been solved for arbitrary Hamiltonians, giving rise to the two implicit solutions \eqref{solX_S} and \eqref{conditionZS}. Finally, by expanding the interaction $V$ over its strength $g$, we have obtained explicit solutions for the minimizations, c.f. Eqs.\ \eqref{eq:finalsolutionX} and \eqref{eq:finalsolutionZ}, which have allowed us to compute the correcting terms due to strong coupling up to order $O(g^3)$, see \eqref{weakcouplingsolution}.

\subsection{Upper bounds and the macroscopic limit}
In this section, we derive simple upper bounds to the strong-coupling correction terms,  $\Delta F^{\rm (irr)}_{\rm min}$ and $\Delta F^{\rm (res)}_{\rm min}$, which in particular imply that they are bounded by $2\norm{V}$
In this section we make no assumptions on the strength of the interaction.
 We begin with the residual free energy $\Delta F^{\rm (res)}$, obtaining
\begin{align}
\Delta F^{\rm (res)}_{\rm min} &=\min_{H_{S}^{(N)}} \Delta F^{\rm (res)} \nonumber\\
&\leq F(\omega_\beta(H^{(0)}+V),H^{(0)})-F(\omega_\beta(H^{(0)}),H^{(0)})
\nonumber\\
&={\rm Tr}(V(\omega_\beta(H^{(0)})-\omega_\beta(H^{(0)})+V))
\nonumber\\
&\quad
-T{\rm S}(\omega_\beta(H^{(0)})\|\omega_\beta(H^{(0)}+V).
\end{align}
In this expression, we have used again the relation $F(\rho,H)-F(\omega_\beta(H),H)=T {\rm S}(\rho \|\omega_\beta(H))$. Since the relative entropy is positive, we obtain
\begin{align}
\Delta F^{\rm (res)}_{\rm min}&\leq {\rm Tr}\left(V(\omega_\beta(H^{(0)})-\omega_\beta(H^{(0)}+V))\right)\nonumber\\
&\leq 2 \|V\|,
\end{align}
which is the desired result.
One can proceed in a similar manner for $\Delta F^{\rm (irr)}_{\rm min}$. First, one realizes that
\begin{align}
&\Delta F^{\rm (irr)}_{\rm min}=  \min_{H_{S}^{(1)}} \Delta F^{\rm (irr)}
\nonumber\\
&\leq F(\omega_\beta(\tilde{H}_S)\otimes \omega_\beta(H_B),\tilde{H})-F(\omega_\beta(H^{\star}),H^{\star})
\end{align}
where  recall that $\tilde{H}_S$ satisfies $\rho_S=\omega_{\beta}(\tilde{H}_S)$. Then, 
\begin{align}
 &\Delta F^{\rm (irr)}_{\rm min} \leq F(\omega_\beta(\tilde{H}_S)\otimes \omega_\beta(H_B),\tilde{H}_{S}+V+H_B)
 \nonumber\\
 &\quad-F(\omega_\beta(\tilde{H}_{S}+V+H_B),\tilde{H}_{S}+V+H_B)\nonumber\\
& \leq {\rm Tr}(V(\omega_\beta(\tilde{H}_S)\otimes \omega_\beta(H_B)-\omega_\beta(\tilde{H}_{S}+V+H_B)))
\nonumber\\
&\leq 2 \|V\|.
\end{align}
This completes the derivation of the upper bounds for $\Delta F^{\rm (irr)}_{\rm min}$ and $\Delta F^{\rm (res)}_{\rm min}$.

\subsection{Limit of arbitrarily strong interactions}
Let us here briefly discuss the limit $g\rightarrow \infty$ of arbitrarily
strong interactions.  To do this, denote by $P_V$ the (projector onto the)
ground-state subspace of the interaction $V$ and denote by $H^{(0)}|_V$ the
operator $H^{(0)}=H^{(0)}_S + H_B$ restricted to the ground-state subspace of $V$ and similarly for $\tilde{H}^{(0)}$. Note that if
$V$ is a local interaction only acting on some finite region, then its
ground-state subspace is highly degenerate. 
In the case of very strong coupling we then obtain
\begin{align}
	\lim_{g\rightarrow \infty} \omega_\beta(\tilde H^{(0)} +gV) = \omega_\beta(\tilde H^{(0)}|_V)\oplus 0,
\end{align}
where the direct sum is over $P_V$. Thus, an arbitrarily strong interaction effectively restricts the Hilbert-space to a sub-space. This has important consequences for the correction terms. 
Consider the irreversible free energy $\Delta F^{\mr{(irr)}}$, defined as
\begin{align}
	\Delta F^{(irr)} = \frac{1}{\beta}S\left(\rho(0)\otimes\omega_\beta(H_B) \| \omega_\beta(H^{1}_S + g V+ H_B)\right).\nonumber
\end{align}
In the limit $g\rightarrow \infty$, the second argument is only supported within the ground-state subspace of $V$, while the support of the first argument is not contained in this subspace. But as is well known, the relative entropy $S(\rho \| \sigma)$ diverges whenever the support of $\rho$ is not contained in the support of $\sigma$. We thus find that
\begin{align}
	\lim_{g\rightarrow \infty} \Delta F^{\mr{(irr)}} = +\infty.
\end{align}
Thus in the limit of arbitrarily strong interactions, the optimal protocol is
to never couple the system to the bath. In this case, work can only be
extracted if the initial state $\rho(0)$ is not \emph{passive} \cite{Pusz1978,Lenard1978}. In
particular, no work can be extracted if $\rho(0) =
\omega_{\beta(0)}(H^{(0)}_S)$ for some $\beta(0)>0$ as is the case in
cyclically working thermal machines.

\section{Heat dissipation}
\label{Sec:Heat}

In thermodynamics, it is often the case that the optimal protocols for one task turns out to be also optimal for others.  Here we apply this logic to show that the previous results can be readily applied in order to minimise heat dissipation and maximise efficiency of heat engines strongly coupled to baths.

In order to study heat dissipation, we focus on (non-cyclic) processes where an initial Hamiltonian $H^{(0)}_S$ is modified to $H^{(N)}_S$ (with $H^{(N)}_S$ fixed and  independent of $N$) by putting it in contact with a strongly interacting bath. More precisely, we consider an initially non-interacting Hamiltonian, $H^{(0)}=H^{(0)}_S+H_B$, an initial state  $\rho^{(0)}=\rho_S \otimes \omega_{\beta}(H_B)$, and  the following family of protocols,
\begin{enumerate}
\item A series of quenches are applied to the local Hamiltonian of $S$, $H_S^{(0)} \rightarrow ... \rightarrow H_S^{(1)}$. After that, the interaction between $S$ and $B$ is turned on. 
\item The Hamiltonian of $S$ is modified while $S$ is in contact with $B$ until $H_S^{(N)}$ is reached, the total Hamiltonian reading $H^{(N)}=H_S^{(N)}+V+H_B$. This amounts to a series of quenches $H_S^{(1)} \mapsto \cdots \mapsto H_S^{(N)}$, each followed by an equilibration to the corresponding thermal state.
\item The interaction between $S$ and $B$ is turned off, so that the final Hamiltonian is $H^{(N+1)}=H_S^{(N)}+H_B$.
\end{enumerate} 
Note that this protocol is essentially the same as the optimal protocol for work extraction but without the last step step, which ensured cyclicity.
Instead, now $H_S^{(N)}$ is fixed, and hence so is the final state of $S$,
\begin{align}
\rho_S^{(N)}=\tr_B \left(\omega_{\beta}(H^{(N)})\right).
\end{align}
As a consequence, the entropy change is also fixed 
\begin{align}
\Delta S=S(\rho_S^{(N)})-S(\rho_S^{(0)}).
\end{align} 
 Then, in the spirit of the second law, our aim is to relate $\Delta S$ with the dissipated heat $Q$, and to find the protocol that minimises $Q$.

Firstly, in order to define the average dissipated heat, we invoke the first law of thermodynamics
\begin{equation}
Q=-\Delta E_S -W.
\label{firstlawapp}
\end{equation}
Since at the beginning and at the end of the process the Hamiltonian is non-interacting, $\Delta E_S$ is simply the change of local energy of the system.
Hence $Q=\Delta E_B$, which corresponds to the energy dissipated to the bath.

Secondly, since $\Delta E_S$ is fixed, it naturally follows that minimizing heat dissipation corresponds to maximizing extractable work. This implies that in the optimal protocol step 2. corresponds to an isothermal process \cite{Gallego14}. In this case, we have  that the total work of steps 1.-3. is
\begin{align}
W = F(\rho^{(0)},H^{(0)}) - F(\omega^{(N)},H^{(0)}) -  \Delta F^{(\rm irr)}.
\end{align}
Hence, using \eqref{firstlawapp},
\begin{align}
Q =& -\Tr(H_S \rho_S^{(N)}) +  \Tr(H_S \rho^{(0)}) - W\nonumber\\
=& -F(\rho^{(0)},H_B)+F(\omega^{(N)},H_B)+T{\rm S}(\rho^{(0)} \|\omega^{(1)})
\nonumber\\
=& -F(\omega_B^{(0)},H_B)+F(\rho^{(N)}_{B},H_B)
\nonumber\\
&+T({\rm S}(\rho_S^{(0)}) +{\rm S}(\rho_B^{(N)}) - {\rm S}(\omega^{(N)})) + \Delta F^{(\rm irr)}.
\end{align}
where we note  $\rho_S^{(N)}=\tr_B(\omega^{(N)})$ and $\rho_B^{(N)}=\tr_S(\omega^{(N)})$.
We now make use of the mutual information to write
\begin{align}
{\rm S}(\omega^{(N)}) = {\rm S}(\rho_{S}^{(N)}) +{\rm S}(\rho_{B}^{(N)}) - I(\omega^{(N)}; S:B) ,
\end{align}
and using again $F(\rho,H)-F(\omega_\beta(H),H)=T {\rm S}(\rho \|\omega_\beta(H))$, we finally obtain 
\begin{align}
\hspace{-0.5mm}Q \hspace{-0.5mm} =\hspace{-0.5mm}  -T \Delta {\rm S}_S
\hspace{-0.5mm}+ \hspace{-0.5mm}T \hspace{-0.5mm}\left(\hspace{-0.5mm}S(\rho^{(N)}_{B}\|\omega_B^{(0)}) + I(\omega^{(N)}; S:B)\hspace{-0.5mm}\right)\hspace{-0.5mm} + \hspace{-0.5mm}\Delta F^{(\rm irr)},
\label{Qstrong}
\end{align}
with $\Delta S_S = S(\rho_S^{(N)})-S(\rho_S^{(0)})$.
As expected, the second law is satisfied in the form $-T\Delta S \geq Q$.  
The protocol minimizing heat dissipation can now be found by minimizing  the positive terms  in \eqref{Qstrong}. Note however that the second term in \eqref{Qstrong} is now fixed through $H_S^{(N)}$. The only task left is hence to minimise $\Delta F^{(\rm irr)}$ over  the Hamiltonian $H_S^{(1)}$, a problem which was already solved yielding \eqref{solX_S}. 

Summarizing, the protocol minimizing heat dissipation in the strong coupling regime --given the initial conditions $H^{(0)}=H^{(0)}_S+H_B$,  $\rho^{(0)}=\rho_S \otimes \omega_{\beta}(H_B)$-- can be described as:  $S$ and $B$ are put in contact through the choice \eqref{solX_S},  which ensures   minimal heat dissipation, and afterwards an isothermal transformation is implemented until the desired Hamiltonian $H_S^{(N)}$ is reached. 

As we did for work extraction, we now replace $V$ by $gV$ and find the first non-vanishing corrections of the penalizing terms in \eqref{Qstrong}. First of all,  to understand the scaling of the term involving the mutual information, let us write the mutual information as
\begin{align}
I(\omega^{(N)};S:B) = {\rm S}\left(\omega^{(N)} \|\tr_B (\omega^{(N)})\otimes \tr_S (\omega^{(N)})\right)\geq 0.
\end{align}
Since we have $\omega^{(N)} = \omega_{S}^{(N)}\otimes
\omega_{B}^{(N)}$
if and only if $g=0$, the function obtains its minimal value $0$ only at $g=0$.
We thus conclude that the corrections in $g$ will be of the order $g^2$ for small $g$. Similarly, as argued in the previous section, the corrections of the relative distances $S(\cdot||\cdot)$ are also of $O(g^2)$. Hence, without doing any explicit calculation, we can already conclude that,
\begin{align}
Q = -T \Delta {\rm S}_S +  K_q g^2 - O(g^3),
\end{align}
where $K_q$ is given by
\begin{align}
 K_q=&\frac{1}{2} \frac{d^2}{dg^2} \bigg[ \Delta F^{(\rm irr)}_{\rm min}+
 \nonumber\\
 &+T\left(S(\rho^{(N)}_{B}\|\omega_B^{(0)}) + I(\omega^{(N)}; S:B)\right) \bigg] \bigg|_{g=0} .
\label{derivativeI}
\end{align}
The computation of $K_q$ can be carried out for each particular model of interest. Here, by using  our previous considerations, we can provide a general and simple lower bound to it,
\begin{align}
K_q \geq \frac{1}{2} \frac{d^2}{dg^2}  \Delta F^{(\rm irr)}_{\rm min}  \bigg|_{g=0} =\frac{1}{2}  \mathrm{cov}_{\omega_\beta(\tilde{H}^{(0)})}(\tilde{V},\tilde{V}).
\label{LowerBoundK_q}
\end{align}
Hence we see that the correction for work extraction in the strong coupling regime also  provides a correction to heat dissipation.

\section{Carnot engine}
\label{Sec:CarnotEngines}

Let us now optimize a Carnot engine in the strong coupling regime, by considering the presence of two baths, $B_c$ and $B_h$, at temperatures $T_c,T_h>0$, respectively. We consider a Carnot cycle consisting of four steps
\begin{enumerate}
\item  \emph{Coupling to $B_h$ + isothermal process+ decoupling from $B_h$}. We consider an isothermal transformation --following the three steps described in  Section  \ref{Sec:Heat} --  from $H^{(A)}_{S}$ to  $H^{(B)}_{S}$. Note that the state after the transformation is,
\begin{align}
\rho_1=  \tr_B \left( \omega_{\beta_h}(H^{(B)}) \right)
\label{reduced1}
\end{align} 
where $H^{(B)}=H^{(B)}_{S}+V_{h}+H_{Bh}$, and $V_h$ corresponds to the interaction $SB_h$ and $H_{Bh}$ to the local Hamiltonian of $B_h$.
\item \emph{Adiabatic expansion.} This is represented by a  quench from $H^{(B)}_{S}$  to $H^{(C)}_{S}$.
\item  \emph{Coupling to $B_c$ + isothermal process+ decoupling from $B_c$}. An isothermal transformation  is applied  from $H^{(A)}_{S}$ to  $H^{(B)}_{S}$. The state after the transformation reads 
\begin{align}
\rho_2 =\tr_B \left( \omega_{\beta_c}(H^{(D)}) \right)
\label{reduced2}
\end{align} 
with $H^{(D)}=H^{(D)}_{S}+V_{c}+H_{Bc}$.
\item Adiabatic compression. A quench back to $H^{(A)}_{S}$ is implemented.
\end{enumerate}
With regard to efficiency, such protocols, which only contain one
isothermal process with each bath per cycle, are optimal. Intuitively, this
follows since every time the working system is coupled to a heat bath, there is
unavoidable dissipation, which decreases the efficiency. It is thus optimal to
couple to each bath the minimal number of times so that work extraction remains
possible. In section~\ref{sec:carnot_optimal} we prove that this intuition is
indeed correct.

In the weak coupling regime, the efficiency of the engine is maximised (obtaining the well-known Carnot bound $T=1 - \beta_h/\beta_c$) when 
\begin{align}
\omega_{\beta_h}(H_S^{(B)})=  \omega_{\beta_c}(H_S^{(C)})
\nonumber\\ 
\omega_{\beta_c}(H_S^{(D)})=  \omega_{\beta_h}(H_S^{(A)})
\label{CarnotWeak}
\end{align}
which ensures no dissipation at any point. Hence,  given two Hamiltonians, e.g. $H_S^{(B)}$ and $H_S^{(D)}$, the engine becomes maximally efficient when the other two Hamiltonians, $H_S^{(A)}$ and $H_S^{(C)}$,  satisfy condition \eqref{CarnotWeak}. We now look for the corresponding condition in the strong coupling regime. We find convenient to  fix $H_S^{(B)}$ and $H_S^{(D)}$ -- so that states $\rho_1$  and $\rho_2$ are fixed -- and optimize over $H_S^{(A)}$ and $H_S^{(C)}$.

The efficiency of the cycle is defined as
\begin{equation}
\eta=-\frac{W}{Q_h},
\end{equation}
the minus sign appearing due to the sign convention. By reasonably assuming that the interaction of $S$ with the cold (hot) bath destroys the correlations of $S$ with the other bath
(or equivalently, use different baths for each round of the cycle),  the initial state before interacting with the hot (cold) bath is   $\rho_1 \otimes \omega_{B_h}$ ($\rho_2 \otimes \omega_{B_c}$), where $\omega_{B_{h/c}}=\omega_{\beta}(H_{B_{h/c}})$. Then we can readily apply  Eq.~\eqref{Qstrong} obtaining  
\begin{align}\label{eq:Qcold}
Q_h= &-T_h\Delta {\rm S} + T_h \bigg(S(\rho^{(B)}_{B_h}\|\omega_{B_h})\nonumber\\
& + I( \omega_{\beta_h}(H^{(B)}); S:B)\bigg) + \Delta F^{(\rm irr)}_h, \nonumber\\
Q_c= &T_c\Delta {\rm S}+ T_c \bigg(S(\rho^{(D)}_{B_c}\|\omega_{B_c})\nonumber\\
& + I( \omega_{\beta_c}(H^{(D)}); S:B)\bigg) + \Delta F^{(\rm irr)}_c,
\end{align}
where $\Delta {\rm S} ={\rm S}(\rho_1)-{\rm S}(\rho_2)$ is the entropy loss of the system $S$, $\rho^{(B)}_{B_h}=\tr_S \omega_{\beta_h}(H^{(B)})$, $\rho^{(D)}_{B_c}=\tr_S \omega_{\beta_c}(H^{(D)})$ and
\begin{align}
 \Delta F^{(\rm irr)}_{h/c}=&F(\rho_{1/2} \otimes \omega_{B_{h/c}},H^{(A/C)})
 \nonumber\\
 &-F(\omega_\beta(H^{(A/C)}),H^{(A/C)}).
\end{align}
The first law of thermodynamics implies that for a cyclic process $W=-Q_h-Q_c$ and
the efficiency becomes
\begin{equation}\label{eq:efficiency-appendix}
\eta= 1+\frac{Q_c}{Q_h}=1-\frac{T_c(1+x_c)}{T_h(1-x_h)},
\end{equation}
where $x_{c/h}$ is the fraction of free energy irreversibly lost during the cold/hot part of the cycle,
\begin{equation}
x_{c/h} = \frac{\Delta F^{\mathrm{(res)}}_{B_{c/h}}+ T_{c/h}I(\omega_\beta(H_{B_{c/h}}); S:B_{c/h})+\Delta F_{c/h}^{\mathrm{(irr)}}}{T_{c/h}\Delta {\rm S} }.
\end{equation}
Now, for an engine producing work, we have that $\Delta E_{B_h}=-Q_h >0$, and hence also $ \Delta {\rm S}>0$, which implies that $x_{c/h}>0$. It then follows that the efficiency is lower than the Carnot efficiency, i.e., $\eta < \eta^{{\rm C}}\coloneqq 1- T_c/T_h$, as expected.

In order to  
maximise the efficiency $\eta$, one needs to minimise both $x_{c/h}$. Crucially, since $H_S^{(B)}$ and $H_S^{(D)}$  are fixed, we only need to minimise the term $\Delta F_{c/h}^{\mathrm{(irr)}}$, which was already minimise leading to \eqref{solX_S}. This implies that the condition \eqref{CarnotWeak} naturally generalises in the strong coupling to
\begin{align}
&\tr_B \left( \omega_{\beta_h}(H^{(B)}) \right)=  \tr_B \left( \omega_{\beta_c}(H^{(C)}) \right),
\nonumber\\ 
&\tr_B \left( \omega_{\beta_c}(H^{(D)}) \right) = \tr_B  \left( \omega_{\beta_h}(H^{(A)}) \right).
\label{CarnotStrong}
\end{align}
These conditions define the optimal choice of $H_S^{(A)}$ and $H_S^{(C)}$  in order to maximise $\eta$ in the strong coupling regime.

In order to study the limit of small interactions for the Carnot engine, we first note that $\Delta {\rm S}$  does depend on $g$, as the initial/final state of S before/after being coupled with the heat bath is the reduced of a thermal state. We can then expand it over $g$ obtaining 
\begin{equation}\label{tappdd}
T_{c,h} \Delta {\rm S} = T_{c,h} \Delta {\rm S}^{\rm (weak)} +K_{S} g+ O(g^2),
\end{equation}
where $\Delta {\rm S}^{\rm (weak)}=S(\omega_{\beta}(H_S^{(B)}))-S(\omega_{\beta}(H_S^{(D)}))$. Then it follows 
\begin{align}
x_{c/h} &= \frac{T_{c/h}K_q^{c/h}g^2+O(g^3)}{T_{c,h} \Delta S^{\rm (weak)} +K_{S} g+ O(g^2)}
\nonumber\\
&= \frac{K_q^{c/h}}{T_{c/h} \Delta S^{\rm (weak)} }g^2+O(g^3),\nonumber
\end{align}
and
\begin{align}
	\frac{1+x_c}{1-x_h}=1+ \left(\frac{K_q^c}{Q_c^{\rm (weak)}} +\frac{K_q^h}{Q_h^{\rm (weak)}}\right)g^2 + O(g^3),
\end{align}
where $Q_{c/h}^{\rm (weak)}=T_{c/h} \Delta S^{\rm (weak)}$. Using \eqref{eq:efficiency-appendix} and the lower bound \eqref{LowerBoundK_q} for $K_q^{c/h}$, we obtain the desired corrections to Carnot 
due to strong coupling.

\subsection{Carnot-like protocols are optimal}
\label{sec:carnot_optimal}
In this section we prove that the Carnot-like protocols considered above are optimal from the point of view of efficiency. 
To do this, let us first discuss some preliminaries. Consider an arbitrary
cyclic protocol. In every such protocol, there are parts of the protocol where
the system remains coupled to one of the heat baths, while multiple quenches
from some Hamiltonian $H$ to some other Hamiltonian $H'$ are done on the
system. The first observation to make is that the protocol can only become more
efficient if we replace this part of the protocol by an isothermal reversible
process from $H$ to $H'$ since such processes are reversible.
This shows that optimal protocols will consist only of two kinds of operations:
\begin{enumerate}
	\item isothermal reversible processes in contact with one of the baths,
	\item quenches while not being coupled to the baths (adiabatic compression).
\end{enumerate}
Any such protocol is thus composed of $n_h$ isothermal process with the hot
bath and $n_c$ isothermal process with the cold bath, with adiabatic quenches
in-between. We can then describe any such protocol by $n_h$ pairs of
Hamiltonians $(H_{h,c}^{(i)},H_{h,c}^{(f)})$, denoting the initial and final
Hamiltonian of the $j$-th IRP with the hot bath, and analogously $n_c$ pairs
of Hamiltonians $(H_{c,j}^{(i)},H_{c,j}^{(f)})$ for the cold bath. Here, we take the convention that
\begin{align}
	H_{c/h,j}^{(i/f)} = H_{c/h,j,S}^{(i/f)} + V + H_B.
\end{align}

Suppose now that a given protocol would have two isothermal processes with one
of the baths after each other, only separated by an adiabatic quench. Then this part of the protocol would in general be irreversible and have fixed initial and final states. For
concreteness, suppose this would happen with the cold bath and suppose the two
initial and final Hamiltonians would be given by
$(H_{c,j}^{(i)},H_{c_j}^{(f)})$ with $j=1,2$. Then we could replace this part
of the protocol by an isothermal reversible process from $H_{c,1}^{(i)}$ to
$H_{c,2}^{(f)}$ and the efficiency could only increase. This shows that in
optimal protocols, the isothermal reversible process at the two different baths
alternate, so that an isothermal process at the hot bath is necessarily
followed by an adiabatic quench and an isothermal process at the cold bath (and
vice-versa).  

Since the protocols have to be cyclic, we already know that $n_h=n_c$. What remains to be shown is that optimal protocols have $n_h=n_c=1$. To see this, suppose, for concreteness, that $n_h=n_c=2$. We will now show that we can always describe such a protocol by two sub-cycles with $n_h=n_c=1$ that are run sequentially. 
To do this, let us start the description of the total protocol at the end of the second isothermal at the cold bath, thus starting with state $\omega_{\beta_c}(H_{c,2}^{(f)})$. Then the original protocol proceeds as follows:
\begin{enumerate}
	\item Adiabatic quench to $H_{h,1}^{(i)}$,
	\item Isothermal process to $H_{h,1}^{(f)}$,
	\item Adiabatic quench to $H_{c,1}^{(i)}$,
	\item Isothermal process to $H_{c,1}^{(f)}$,
	\item Adiabatic quench to $H_{h,2}^{(i)}$,
	\item Isothermal process to $H_{h,2}^{(f)}$,
	\item Adiabatich quench to $H_{c,2}^{(i)}$,
	\item Isothermal process back to $H_{c,2}^{(f)}$.
\end{enumerate}
We have here omitted turning on and off the interaction between the system and bath. Let us now replace this protocol by the following protocol, which has exactly the same efficiency:
\begin{enumerate}
	\item Adiabatic quench to $H_{h,1}^{(i)}$,
	
	\item Isothermal process to $H_{h,2}^{(f)}$ (change here),
	\item Isothermal process from $H_{h,2}^{(f)}$ to $H_{h,1}^{(f)}$,
	\item Adiabatic quench to $H_{c,1}^{(i)}$,
	\item Isothermal process to $H_{c,1}^{(f)}$,
	\item Adiabatic quench to $H_{h,2}^{(i)}$,
	\item Isothermal process to $H_{h,1}^{(f)}$ (change here),
	
	\item Isothermal process from $H_{h,1}^{(f)}$ to $H_{h,2}^{(f)}$.
	\item Adiabatic quench to $H_{c,2}^{(i)}$,
	\item Isothermal process back to $H_{c,2}^{(f)}$.
\end{enumerate}
The only change that occurred is a splitting off of isothermal processes into
two pieces. Due to reversibility of isothermal processes, this does not change
the work and heat flows and hence this protocol has the same efficiency.
However, we can now see that we have turned the protocol into two simple
cycles: One composed of one isothermal in contact with the hot bath from
Hamiltonian $H_{h,1}^{(i)}$ to $H_{h,2}^{(f)}$ and an isothermal 
with the cold bath connecting the Hamiltonians $H_{c,2}^{(i/f)}$. The second
cycles consists of the isothermal process in contact with the hot bath connecting
the Hamiltonians $H_{h,2}^{(i)}$ and $H_{h,1}^{(f)}$ and an isothermal with the
cold bath connecting the Hamiltonians $H_{c,1}^{(i/f)}$ as before. The two
cycles are connected by one isothermal from $H_{h,1}^{(f)}$ to $H_{h,2}^{(i)}$
and the same isothermal run backwards. Due two reversibility, these two
isothermals "cancel out" when calculating heat and work. We thus conclude that
the total work of the protocol and the total heat absorbed from the hot baths
are given by
\begin{equation}
\begin{split}
W &= W_1 + W_2\, , \\
Q &= Q_1 + Q_2\, ,
\end{split}
\end{equation}

where $W_1$ and $W_2$ denote the work in the individual cycles and similarly for the heat. The total efficiency then yields
\begin{align}
	\eta  = \frac{W_1+W_2}{Q_1+Q_2} \leq \max\left \{\frac{W_1}{Q_1},\frac{W_2}{Q_2}\right\}. 
\end{align}
A similar construction can be made for any protocol with $n_h=n_c>1$. Hence, the efficiency of any such protocol can be bounded as 
\begin{align}
	\eta \leq \max\{\eta_j\},
\end{align}
where $\eta_j$ dentoes the efficiency of the $j$-th sub-cycle. 
We conclude that optimal protocols are simple Carnot-cycles as analyzed in the previous sections.

\section{Power and lower bound on the equilibration time}\label{sec:power-and-time}
\label{Sec:EquilibrationTime}
We have argued in the main text that a dimensional analysis suggests that the
equilibration time should scales as $1/g$.
In this section, we show that the equilibration time satisfies $\tau \geq C/g$ for some constant
$C>0$.
By equilibration we mean the process in which the initial expectation value of
any operator $A(0)$ evolves in time towards a certain value $\bar A$
in which it (approximately) remains.

\subsection{Lower bound on the equilibration time for a single equilibration}

In order to give a lower bound for the equilibration time let us
consider the fastest
rate of change of $\average{A(t)}$, i.e., the quantity
\begin{equation}
v=\sup_t\left|\frac{\mathrm{d}}{\mathrm{d} t}
  \average{A(t)}\right|
\end{equation}
which is trivially upper bounded by
\begin{equation}
v = \sup_t\left|\tr\big([A(t),H]\rho\big)\right|
\leq \sup_t \norm{[A(t),H]} = \norm{[A,H]}.
\end{equation}
This upper bound implies a lower bound on the equilibration time 
by means of 
\begin{equation}
\tau v \geq |A(0)-\bar A|\, .
\end{equation}
In the particular case that $A=H_S$, the rate at which the energy of the system
changes during the equilibration is bounded by
\begin{equation}
v \le  \norm{[H_S,H]}=g\norm{[H_S,V]}
\end{equation}
with $c=\norm{[H_S,V]}$.
This leads to an equilibration time lower bounded by
\begin{equation}\label{eq:bound-single-equilibration}
\tau \geq \frac{|\Delta E_S|}{g c}.
\end{equation}

\subsection{Lower bound on the equilibration time for the entire cycle of the heat engine}

In order to lower bound the equilibration time of the entire cycle,
it will be useful to use the bound of the equilibration time of a
single equilibration by means of the energy change of the bath
\begin{equation}\label{eq:bound-single-equilibrationII}
\tau
\geq \frac{\delta E_B}{g r},
\end{equation}
where now $r\coloneqq \norm{[H_B,V]}$.
Although in general $r$ could scale as $\norm{H_B}$, in practice the Hamiltonian of the
bath has a locality structure and the commutator is only non-trivial on the degrees of freedom
close to the boundary and $r$ is independent of the bath's size.
Let us decompose the cycle in a heat engine described above into two main parts.
\begin{itemize}
\item \emph{Coupling to the cold bath + isothermal reversible process with it + decoupling from it}.
The system is initially already decoupled from the hot bath and thus the global Hamiltonian is non interacting. After performing the isothermal process with the cold bath, the system is also decoupled from the bath. Hence, the accumulated energy variation of the bath during all the protocol steps is given by
\begin{align}
\sum_i |\delta E_{B_c}^{(i)}|\geq \left|\sum_i \delta E_{B_c}^{(i)}\right|=|\Delta E_{B_c}|=|Q_c| .
\end{align}
A lower bound on the time that such part of the protocol requires is given by
\begin{align}
\Delta t_c\geq \sum_i \frac{|\delta E_{B_c}^{(i)}|}{gr_c} \geq \frac{|Q_c|}{gr_c}
\end{align}
where $r_c= \norm{[H_{B_c},V_c]}$ is the maximum rate at which the bath loses or gains energy, $H_{B_c}$ is the Hamiltonian of the cold bath, and $V_c$ is the interaction that couples the system to the cold bath.

\item \emph{Coupling to the hot bath + isothermal reversible process with it + decoupling from it}.
By means of exactly the same argument, the time required to run the second part of the cycle
can be lower bounded by
\begin{equation}
\Delta t_h \geq \frac{|Q_h|}{gr_h},
\end{equation}
where $r_h= \norm{[H_{B_h},V_h]}$, $H_{B_h}$ is the Hamiltonian of the hot bath, and $V_h$ is the interaction that couples the system to the hot bath.
\end{itemize}

The total run-time of the cycle is then bounded by
\begin{align}\label{eq:bound-run-time-cycle}
\Delta t &=\Delta t_c +\Delta t_h \geq \frac{|Q_c|}{gr_c}+\frac{|Q_h|}{gr_h} \nonumber\\&=
|Q_h|\left(\frac{1}{gr_c}+\frac{1}{gr_h}\right)-\frac{|W|}{gr_c},
\end{align}
where we have considered that for an engine $|Q_c|=|Q_h| - |W|$.

\subsection{Upper bound on the power of the heat engine}

From equation \eqref{eq:bound-run-time-cycle} we obtain a limit in the power of a heat engine
in terms of its efficiency and coupling strength, that is,
\begin{equation}
P\coloneqq \frac{W}{\Delta t}\le \frac{gr_c \eta}{1-\eta +\frac{r_c}{r_h}}< gr_h \eta \, ,
\end{equation}
where we have used the definition of efficiency $\eta=W/Q_h$ (with $W,Q_h>0$ for an engine) and the fact that $1-\eta >0$.

\section{ Caldeira-Leggett model}
\label{Sec:CLmodel}
In this section, we provide more details about the Caldeira-Leggett (CL) or Ullersma
model and the exact spectral density that we are using.
The CL-model describes a central harmonic oscillator (the system) coupled to $ N $ peripheral modes (constituting
the bath), so that the Hamiltonian takes the form
\begin{equation}
H=H_S+H_B+gV+H_R
\label{HamiltonianCL}
\end{equation}
for which
\begin{align}
H_S&= \frac{1}{2}\left(m\omega^2 x^2 +\frac{p^2}{m}\right),\\
H_B&= \frac{1}{2}\sum_{\mu}\left(m_\mu\omega_\mu^2 x_\mu^2+\frac{p_\mu^2}{ m_\mu}\right),
\label{HCLB}\\
V&=x\sum_{\mu}g_\mu x_\mu,
\label{HCLV}\\
H_L&= x^2 g^2 \sum_{\mu} \frac{g_\mu^2}{m_\mu\omega_\mu^2},
\label{HCLL}
\end{align}
here, the coordinates $ \{x,p\} $ refer to the system $S$ and $ \{x_{\mu},p_{\mu}\} $ to the bath oscillators. 
$ H_L $ is the frequency-shift  (Lamb-shift) needed to compensate for the distortion induced by the coupling term on the effective potential of the central oscillator. We also note that we introduced the parameter $g$ at hand (see \eqref{HamiltonianCL} and \eqref{HCLL}), in order to quantify the strength of the system-bath interaction.
The dynamics of $S$ depends only on the spectral density of the bath $J$, which is defined as
\begin{equation}
J(\omega) := \frac{\pi}{2} \sum_\mu \frac{g_\mu^2}{\omega_\mu}\delta(\omega-\omega_\mu).
\end{equation}
In the continuum limit, the spectral density is often assumed to be well approximated by a continuous
function. A common choice of $J$ is the so called Ohmic spectral density that takes the form
\begin{equation}
J(w)=  \eta \omega
\label{ohmicJ(w)}
\end{equation}
for frequencies $\omega$ significantly smaller than some cut-off $\Omega>0$.
In our work we are interested in large but finite $n$. In order to discretise the above considerations, we assume that the bath frequencies are  uniformly distributed,
\begin{equation}
\omega_\mu = \frac{\mu}{n} \Omega,   
\label{omega_k}
\end{equation}
$\mu=1,\dots,n$, where $\Omega$ is the highest frequency, and then,
\begin{align}
\frac{\pi}{2} \sum_{\mu=1}^{n} \frac{g_\mu^2}{\omega_\mu}\delta(\omega-\omega_\mu) \approx \frac{n}{\Omega} \frac{\pi}{2}  \int_{0}^{\Omega}  \frac{g(x)^2}{x}\delta(\omega-x) dx 
\end{align}
where we note that $dx \sim  \Omega/n$. From \eqref{ohmicJ(w)}, we obtain,
\begin{align}
\eta \omega \approx \frac{ g(\omega )^2n}{\Omega \omega}.
\end{align}
Going back to the discrete indices, and with $\eta=1$, we finally obtain
\begin{equation}
g_{\mu} = \omega_{\mu} \sqrt{\frac{2 \Omega}{\pi n}}.
\label{J(w_k)}
\end{equation}
 Together with the assumption $m_\mu =1$, the relations  \eqref{omega_k} and \eqref{J(w_k)} fully determine the Hamiltonian \eqref{HamiltonianCL}. In fact, the only relevant variable that we consider is the interaction strength $g$ in \eqref{HamiltonianCL}.

\begin{figure}
  \includegraphics[width=0.9\columnwidth]{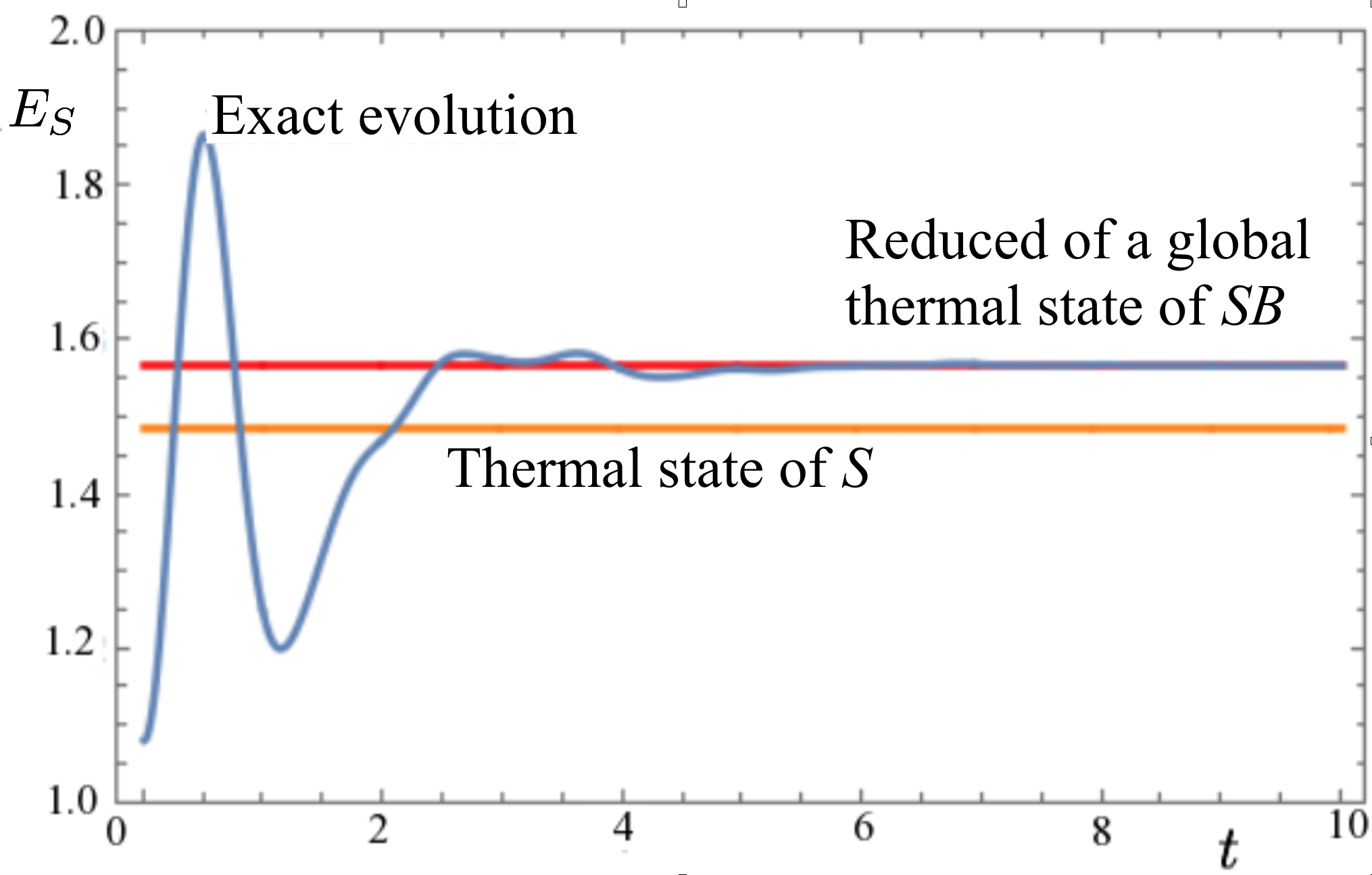}

   \caption{Exact time evolution of the energy expectation 
   of an quantum harmonic oscillator $S$ with frequency $\omega=1$ interacting strongly ($g=1$) with a thermal bath according to the Caldeira-Leggett-model.
 The thermal bath consists of $50$ harmonic oscillators with $m_k's=1$ and equally distributed frequencies in the range $(0,5\omega)$. For $S$ we have $m=1$ and $\omega=1$, and it is set initially in a Gibbs state at temperature $\beta_S=1$, whereas $B$ is at  $\beta=0.7$.
}
\label{EquilibrationCaldeiraLeggett}
\end{figure}

Let us now briefly outline how to solve quadratic bosonic systems exactly, which we apply to the Caldeira Leggett model. We refer the reader to 
Refs.\ \cite{Rivas2010,AreaReviewE} for more detailed and extensive derivations and explanations. 
First, we define the vector of canonical coordinates as
${\bf r}=(x,x_1,x_2,\dots,x_n,p,p_1,\dots,p_n)^T$.
Then we can express the total Hamiltonian \eqref{HamiltonianCL} as
\begin{align}
H=\frac{1}{2}    {\bf r}^{\dagger} H_r      {\bf r}.
\end{align}
Now we invoke Williamson's theorem to write
\begin{align}
H_r= S^{\dagger} (D \oplus D) S
\end{align}
where $D={\rm diag} (d_1,\dots,d_{n+1})$ and the main diagonal elements are given by the 
strictly positive square roots of the spectrum of $(i\sigma H)^2$. The total Hamiltonian can then be written as
\begin{align}
H= \frac{1}{2} {\bf q}^{\dagger} (D\oplus D) {\bf q}
\end{align}
with ${\bf q}= S {\bf r}$. This expression gives rise to
\begin{align}
H= \sum_{k} d_k \biggl(b_k b_k^{\dagger}+\frac{1}{2}\biggr)
\end{align}
where ${\bf b}= \Omega^{-1} S {\bf r}$, and
 \begin{align}
\Omega &= \frac{1}{\sqrt{2}} \begin{bmatrix}
           \mathbb{I} & \mathbb{I}  \\
           -i\mathbb{I} & i\mathbb{I}  \\
         \end{bmatrix}.
  \end{align}
Let us now define the first and second moments of $\rho :=  \rho$ as
${\bf m}_i(\rho) = \tr(\rho r_i)$, and 
\begin{align}
  \gamma_{i,j}(\rho) &=\tr \left(\rho (r_i r_j+r_j r_i) \right)-2m_i m_j.
\end{align}
Given these definitions, it can be shown that the time evolution of 
${\bf m}(\rho(t))$ and $\gamma(\rho(t))$ under $H$, with $\rho(t)=e^{-iHt}\rho e^{iHt}$,
\begin{align}
{\bf m}(\rho(t)) &= e^{-\sigma H_r t}  {\bf m}(\rho)
\nonumber\\
\gamma_{i,j}(\rho)&=\left( e^{-\sigma H_r t} \gamma(\rho) e^{H_r\sigma t} \right)_{i,j}
\label{gammatt}
\end{align}
where
 \begin{align}
\sigma &=  \begin{bmatrix}
           0 & -\mathbb{I}  \\
           \mathbb{I} & 0 \\
         \end{bmatrix}.
  \end{align}
  Through the exact evolution \eqref{gammatt}, one can compute, e.g., the energy as a function of time as in Fig.\ \ref{EquilibrationCaldeiraLeggett}.
  
\subsection{Equilibration in the Caldeira-Leggett model}
\label{sec:app:equilibration}
\begin{figure}[t!b]
  \includegraphics[width=.9\columnwidth]{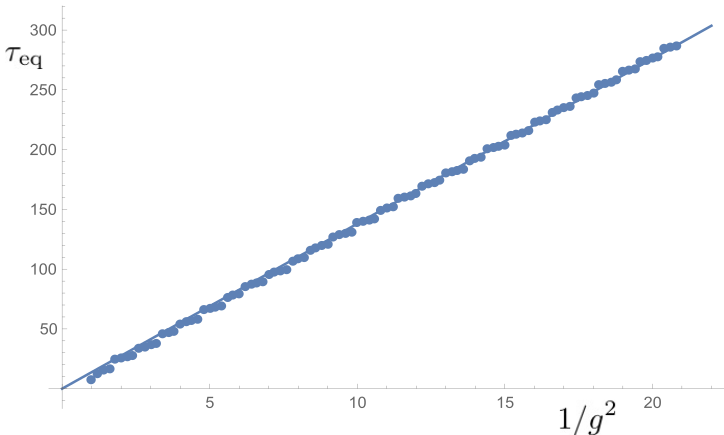}
   \caption{Time of equilibration vs $1/g^2$ in the Caldeira-Leggett model. We take a bath of $n=300$ oscillators with equi-distributed frequencies up to $\Omega=2.1$. As an initial state we take $\rho^{(0)} = \omega_\beta(H_S) \otimes \omega_\beta(H_B)$, with $\omega_S=1$, $\beta_S=1$, $\beta_{B}=3.5$. In order to determine the equilibration time, we let $SB$ evolve the energy of $S$ stays into a region $(0.99a,1.01a)$, for some value $a$.}
\label{PictureTimeEquilibration}
\end{figure}

In this section, we aim at understanding under what conditions the expectation value of
the operator $A$ equilibrates and how long such equilibration process takes.
To do so, we make use of the arguments put forward in Refs.\ \cite{ArnauEquilibration,BerlinEquilibration} in
the Hilbert space, but here in the space of modes.
The time evolution of an observable quadratic in the canonical coordinates
in the Caldeira-Leggett model,
\begin{equation}
A=\sum_{i,j}A_{i,j}r_i r_j\, ,
\end{equation}
for an initial state the covariance matrix of which has entries
$\gamma_{i,j}(0)=\Tr\left((r_i r_j+r_j r_i)\rho(0)\right)$ reads
\begin{equation}
A(t)=\Tr(A\rho(t))=\sum_{k,l}\tilde A_{k ,l}C_{k,l}\e^{\iu (\tilde d_k + \tilde d_{l})t}\, ,
\end{equation}
where $\tilde A = \Omega^T Q \Omega=\Omega^T\sigma S \sigma A \sigma S^T \sigma\Omega$ is a matrix
associated to the observable, $C_{k , l}=\Omega^{-1}(S \gamma^{\bf r}S^T -\iu\sigma) (\Omega^{-1})^T /2$ is the covariance matrix of the initial state,
$\tilde d_k= d_k$ if  $k\le L$ and $\tilde d_k=-d_{k-L}$ for $k>L$, and
\begin{equation}
\Omega = \frac{1}{\sqrt{2}}\left(\begin{array}{cc}\id & \id \\ -\iu\id & \iu\id\end{array}\right)\, .
\end{equation}
If the observable equilibrates, its equilibrium value is the infinite time average
\begin{equation}
\bar A = \lim_{T\to\infty}\frac{1}{T}\int_0^T A(t)\, .
\end{equation}
Let us restrict us for simplicity to the generic case in which the spectrum of $D$
has $\tilde d_k= -\tilde d_{l}$ if and only if $l= k+L \mod L$.
The equilibrium value of $A$ in such a situation reads
\begin{equation}
\bar A= \sum_{k}\tilde A_{k , k+L}C_{k,k+L}
\end{equation}
where the sum in the subindices is taken modulo $L$ and we have identified
a Kronecker delta.

Let us now introduce the \emph{time signal} of an observable $A$ for
the initial state $\rho(0)$ as the distance from the equilibrium
value of the instantaneous expectation value of $A$ at time $t$
\begin{equation}
f(t):=A(t)-\bar A \,,
\end{equation}
and for in absence of degeneracies
\begin{align}
f(t)&=\sum_{k,l}\tilde A_{k ,l}C_{k , l}\e^{\iu (\tilde d_k + \tilde d_{l})t}- \sum_{k}\tilde A_{k , k+L}C_{k,k+L}\nonumber\\
&=\sum_{l\neq k+L}\tilde A_{k , l}C_{k , l}\e^{\iu (\tilde d_k + \tilde d_{l})t}.
\end{align}
It is useful to write the time signal as
\begin{equation}\label{eq:time-signal}
f(t)=\sum_{\alpha}v_{\alpha}\e^{-\iu \omega_{\alpha}t}\, ,
\end{equation}
where  $\omega_\alpha=\omega_{(k,l)}:=\tilde d_k +\tilde d_{l}$ with $k\le l$ and $l\ne k+L$, which in general forms a set of the $2L^2$ different frequencies,
and 
\begin{equation}
v_\alpha=v_{(k,l)}:=\tilde A_{k, l}C_{k, l}+\tilde A_{l,k}C_{l,k} 
\end{equation}
is the relevance of each one.
The restrictions of the sums over $k$ and $l$ are due to the fact $\tilde d_k + \tilde d_{l}=\tilde d_{l} + \tilde d_{k}$ and the subtraction of the equilibrium value $\bar A$.
In this new form \eqref{eq:time-signal}, the time signal can be seen as the sum of a cloud of points in the complex plane which are initially in the position $v_\alpha$ and rotate at 
angular velocity $\omega_\alpha$.

In the same spirit of the works of equilibration in closed quantum systems
and in order to define a notion of equilibration we compute the average distance from equilibrium
\begin{equation}\label{eq:average-distance-from-equilibrium}
\langle|f(t)|^2\rangle_t:=\lim_{T\to\infty}\frac{1}{T}\int_0^T |f(t)|^2 =
\sum_\alpha |v_\alpha|^2
\end{equation}
where to simplify the calculations we have assumed that
the spectrum of $D$ is generic such that $\omega_\alpha=\omega_{\alpha'}$ if and only if $\alpha = \alpha'$.

The average in time of the signal $f(t)$ gives as a notion of to which extent the observable $A$ equilibrates. If $\langle|f(t)|^2\rangle_t\ll 1$, then the observable takes for most of times the equilibrium value. In contrast, if  $\langle|f(t)|^2\rangle_t \simeq O(f(0))$ then the system is most of times out of equilibrium.
Here we assume that the system equilibrates and hence
\begin{equation}\label{eq:equilibration-condition}
\langle|f(t)|^2\rangle_t\ll |f(0)|\, .
\end{equation}
The above condition \eqref{eq:equilibration-condition} implies some type of synchronization of the initial phases of the complex numbers $v_\alpha$.
In particular, if the phases of $v_\alpha$ were isotropically distributed, then
the value of $g(0)\simeq \langle g(t)\rangle_t$ instead of \eqref{eq:equilibration-condition}.
To see this, let $v_{\alpha}=|v_{\alpha}|\e^{\iu\theta_{\alpha}}$
with $\alpha=1,\ldots,d$ be a set of $d$ independent random complex variables
with an isotropic probability distribution $p_\alpha(r,\theta)=p_\alpha(r)=\delta(r-r_\alpha)$, i.~e.\ the random variable $v_\alpha$ has fixed modulus $r_\alpha$ and a
random phase $\theta_\alpha$.
Then, the variance of the random variable is given by
\begin{equation}
\var \left(\sum_{\alpha}v_{\alpha}\right)=\sum_\alpha \var(v_\alpha)
=\sum_\alpha <|v_\alpha|^2>=\sum_\alpha |v_\alpha|^2,
\end{equation}
where we have used the fact that the variance of a sum of independent random variables is the sum of variances and 
the first moments $<v_\alpha>=0$.
In other words, if the phases are random, the typical value of $f(0)=\sum_\alpha v_\alpha$ will be of the order 
\begin{equation}
\left(\sum_\alpha |v_\alpha|^2\right)^{1/2}=
\left(\sum_\alpha |v_\alpha|^2\right)^{1/2}=\left(\langle f(t)\rangle_t\right)^{1/2}.
\end{equation}
Thus, the relaxation to equilibrium has to be understood as the dephasing process of the set of points $v_\alpha$ in the complex plane. Initially, the points $v_\alpha$ are ``more or less'' synchronized in phase, as time runs, they separate each other due to their different angular velocities $\omega_\alpha$. Once they have completely dephased and have formed an isotropic cloud,
the system is at equilibrium.
As argued in Ref.~\cite{ArnauEquilibration}, the estimate of the equilibration time $\tau$ is the inverse of the dispersion of the relevant angular velocities $\omega_k$, that is,
\begin{equation}\label{eq:equilibration-time-scales}
\tau\simeq \Delta \omega^{-1}
\end{equation}
with
\begin{equation}
\Delta \omega^{2}=\sum_{\alpha}p_{\alpha}\omega_{\alpha}^{2}-\left(\sum_{\alpha}p_{\alpha}\omega_{\alpha}\right)^{2}\label{eq:conjecture}
\end{equation}
where the relevance $p_\alpha=|v_\alpha|^2/\sum_{\alpha'}/|v_\alpha|^2$
is the normalized relevance of the frequency $\omega_\alpha$.
In order to understand the behavior of the equilibration time with the strength of the interaction $g$, we need to study how the $\Delta \omega$, and specifically the matrix-elements $|\tilde{A}_{k, l}|$ and $|C_{k, l}|$, change with $g$.
In particular, we study the scaling of their dispersion in $\omega$ of $|\tilde{A}_{k, l}|$ and $|C_{k, l}|$ for different $g$'s in the Caldeira-Leggett model taking $A=H_S$ and find that
both scale as $g^2$.
This together with Eq.~\eqref{eq:equilibration-time-scales} sets
a time-scale which behaves as
\begin{equation}
\tau\simeq g^{-2}\, .
\end{equation}
This is numerically confirmed in Fig.~\ref{PictureTimeEquilibration}, where the
time of equilibration is plot with respect to $1/g^2$ in the Caldeira-Leggett model.
This supports the idea that the underlying mechanism of equilibration in integrable models is also dephasing.

%

\end{document}